\documentclass[11pt]{article}

\usepackage{amsmath,amssymb}
\usepackage{slashed}
\usepackage{xcolor} 
\usepackage{graphicx}
\usepackage{dcolumn} 
\usepackage{bm} 
\usepackage{epsfig}
\usepackage{epstopdf}
\usepackage{grffile}
\usepackage{color}
\usepackage{colordvi}
\usepackage{amsmath,amssymb}
\usepackage{rotating}
\usepackage{lscape}
\usepackage{cite}
\usepackage{float}
\usepackage{hyperref}

\def\Re{{\cal R \mskip-4mu \lower.1ex \hbox{\it e}\,}}
\def\Im{{\cal I \mskip-5mu \lower.1ex \hbox{\it m}\,}}

\def\tev{\,{\ifmmode\mathrm {TeV}\else TeV\fi}}
\def\gev{\,{\ifmmode\mathrm {GeV}\else GeV\fi}}
\def\mev{\,{\ifmmode\mathrm {MeV}\else MeV\fi}}

\begin{document}

\begin{center}

\vspace*{15mm}
\vspace{1cm}
{\Large \bf Study of top quark dipole interactions in $t\bar{t}$ production associated with two heavy 
gauge bosons at the LHC}

\vspace{1cm}

{\bf  Seyed Mohsen Etesami, Sara Khatibi,  and Mojtaba Mohammadi Najafabadi }

 \vspace*{0.5cm}

{\small\sl School of Particles and Accelerators, Institute for Research in Fundamental Sciences (IPM) P.O. Box 19395-5531, Tehran, Iran } \\

\vspace*{.2cm}
\end{center}

\vspace*{10mm}

%
%
\begin{abstract}\label{abstract}
In this paper,  we investigate the prospects of measuring the 
strong and weak dipole moments of the top quark at the Large Hadron Collider (LHC).
Measurements of these couplings provide an excellent opportunity to probe new physics
interactions as they have quite small magnitudes in the Standard Model. Our analyses are
through studying the production cross sections of $t\bar{t}WW$ and  $t\bar{t}ZZ$ processes in the same sign dilepton and four-lepton final states, respectively.
The sensitivities to strong and weak top quark dipole interactions at the $95\%$ confidence level 
for various integrated luminosity scenarios are derived and compared with other studies. 
In addition to using the total cross sections, a novel handle based on an angular observable is introduced which is found
to be sensitive to variations of the top quark strong dipole moments. We also investigate the sensitivity of the 
invariant mass of the system to the strong and weak dipole moments of the top quark.
\end{abstract}

\vspace*{3mm}

PACS Numbers:  13.66.-a, 14.65.Ha

{\bf Keywords}: Top quark, Dipole moments, LHC.

\newpage


\section{Introduction}\label{Introduction}

The search for new physics beyond the Standard Model (SM)
 is one of the main purposes of the CERN Large Hadron Collider (LHC).
 At the LHC, the impacts of beyond the SM physics could be directly seen,
 providing that the characteristic scale would be below the center of mass
 energy of the related hard processes. If not, the new physics effects need to be explored
via the accurate measurements of the couplings of the SM particles.
According to the recent  LHC results,  all measurements are found to be in agreement with the 
SM predictions \cite{r0}. This could be a hint that possible new degrees of freedom are separated in mass from the SM fields.
As a result,  the available energy in the LHC collisions is not enough for
direct production of the heavy degrees of freedom coming from beyond the SM. 
Therefore, one could parameterize the effects of all new physics by a series of $SU(3)_{c}\times SU(2)_{L}\times U(1)_{Y}$
gauge invariant operators $\mathcal{O}_{i}$ constructed out of the SM fields \cite{r1,r2,r3,r4}.
These operators should be of dimension $d > 4$ and typically the
leading effects for collider observables show up at $d = 6$.
Their coefficients are suppressed by inverse powers of the
scale of new physics $\Lambda$:
\begin{eqnarray}
\mathcal{L}_{eff} = \mathcal{L}_{\rm SM} + \sum_{i}\frac{c_{i}\mathcal{O}^{(6)}_{i}}{\Lambda^{2}},
\end{eqnarray}
where $\mathcal{L}_{\rm SM}$ denotes the SM Lagrangian and $c_{i}$ are the dimensionless Wilson
coefficients. Such a {\it model independent} parameterization has the possibility to be linked to 
ultraviolet completions and the results could be interpreted in various beyond the SM theories.
 The dimension six operators $\mathcal{O}^{(6)}_{i}$ have
been listed in Refs.\cite{r1,r2,r3}. Studies on the validity of the SM effective field theory (EFT)
and the fact that the EFT validity range could not be obtained only
on the basis of  low energy information are available in Ref.\cite{r5}.

From the theoretical point of view, top quark could
provide a unique way into beyond the SM physics, since the top quark Yukawa coupling
is the largest among all other SM fermions. 
Particularly, the CP properties of top quark interactions with the SM fields
 is one of the important subjects to study in the top quark sector \cite{r6}.
 Especially, it has been shown that the CP violating couplings of the top quark
 in the framework of a model with an extended scalar sector can
 explain the observed baryon asymmetry of the universe \cite{r7}.
 In the top quark sector, within a beyond SM theory, CP violating interactions  may also 
 show up through the form of electric, strong, and weak dipole moments.
So far, there have been many studies of the potential for revealing possible effects of new physics 
in the top quark sector at the LHC, Tevatron and future colliders 
using the higher-dimensional operators \cite{rsavi,r8,r9,r10,r11,r12,r13,r14,r15,r16,r17,r18,r19,r20,r21,r22,r23,r24,r25,r26,r27,r28,r29,r30,r31,r32,r33,r34,r35,r36,r37,r38,r39,
r40,r41,r42,r43,r44,r45,r46,r47,r48,r49,r50,r51,r52,r53,r54,r55,r550}.

With the LHC phase II upgrade, in which a large amount of data is going to be collected and 
several experimental efforts are going on to assess and reduce the systematic uncertainties, 
rare SM processes will become accessible \cite{r56,r57,r58}. In particular, final states containing several 
heavy SM degrees of freedom could be observed and new physics effects can be studied 
through them as they suffer from small amount of backgrounds.
For instance, $pp\rightarrow t\bar{t} VV$ processes, where $V=Z,W^{\pm}$ are of the
promising channels through which new physics beyond the SM can be investigated.
Studying these processes has the advantage of having naturally high multiplicity
final states and consequently the backgrounds are better under control.
Large thresholds of $2(m_{t}+m_{V})$ with $V=W,Z$ for $t\bar{t}WW$ and $t\bar{t}ZZ$ productions,   
restricts the phase space and lead to small production cross sections at the level
of few femto-barns. However,  LHC is able to reach the threshold and its experiments
are able to observe these processes as around 75 $t\bar{t}ZZ$ and 420 $t\bar{t}WW$ events
are expected to be produced per 30 fb$^{-1}$ of integrated luminosity of data \cite{r56}.
It is worth mentioning that so far the ATLAS and CMS experiments have
measured the top pair production cross sections in association with a single 
W or Z boson \cite{r59,r60}.
Measuring the $t\bar{t}WW$ and $t\bar{t}ZZ$ rates at the LHC is in particular remarkable in top
quark sector as they provide the possibility to probe the top quark couplings with the SM heavy gauge bosons
and even multi-gauge boson interactions. This allows direct probes for dynamics of electroweak symmetry breaking.

In this paper, our concentration is specially on studying the strong and weak electric and magnetic
dipole moments of the top quark through the $pp\rightarrow t\bar{t} WW$ and
$pp\rightarrow t\bar{t} ZZ$ processes at the LHC.
In the SM framework at tree level, 
the magnetic and electric dipole moments are zero and they could be generated at higher order electroweak corrections
which have small magnitudes \cite{r61}.
However, sizable enhancements are predicted in various extensions of the SM \cite{r30,r61}. Therefore, observation of
these moments with deviations from the SM predictions would be indicative of beyond the SM physics.
A highly motivated task would be to investigate how precise these dipole moments can be measured at the collider experiments.

The plan of this paper is as follows. In Section \ref{effL}, the top quark strong and weak dipole moments 
are defined in the context of the SM effective field theory and the relations of the dipole moments with the dimension-six
operators are  given.
Section \ref{sns} is dedicated to estimate the sensitivity 
of the $pp\rightarrow t\bar{t} WW$ and
$pp\rightarrow t\bar{t} ZZ$ processes to the top quark dipole moments and prospects arising from the
production rates.  Section  \ref{ang}  concentrates  on  introducing sensitive observables to the top quark dipole moments.
The conclusions and results  are
summarized  in Section \ref{summary}.

\section{Top quark effective couplings}\label{effL}

As we have mentioned in the previous section, within the SM effective framework, 
the effects of new physics  can be parameterized by using 
higher-dimensional operators involving the SM fields, assuming they come
from new degrees of freedom occurring at a large energy scale $\Lambda$. 
Considering dimension-six operators and following Ref.\cite{r2}, we present 
the general expressions for the gluon-top-antitop ($gt\bar{t}$) and $Z$-top-antitop ($Zt\bar{t}$)
vertices here.

\subsection{$gt\bar{t}$ vertex}
The most general $gt\bar{t}$ coupling considering dimension-six operators including the SM part 
could be parameterized as follows \cite{r2}:
\begin{eqnarray}\label{e1}
\mathcal{L}_{gt\bar{t}} = -g_{s}\bar{t}\frac{\lambda^{a}}{2}\gamma^{\mu}tG^{a}_{\mu}-g_{s}\bar{t}\frac{\lambda^{a}}{2}\frac{i\sigma^{\mu\nu}}{m_{t}}
(d_{V}^{g}+id_{A}^{g}\gamma_{5})tG^{a}_{\mu\nu},
\end{eqnarray}
where $g_{s}$ denotes the strong interaction coupling, $d_{V}^{g}$ and $d_{A}^{g}$ are real parameters
which are related to the top quark chromomagnetic and chromoelectric dipole moments, respectively.
Gell-Mann matrices are denoted by $\lambda^{a}$ and $G^{a}_{\mu\nu}$ is the strong field strength tensor.
At leading-order, in the SM context,  $d_{V}^{g}$ and $d_{A}^{g}$ are zero. 
The first term in Eq.\ref{e1} is the SM interaction, second and third terms which 
consist of both $gt\bar{t}$ interaction and four-leg $ggt\bar{t}$ coupling come from the dimension six operator \cite{r2}:
\begin{eqnarray}
 O^{33}_{uG\phi} \sim (\bar{q}_{L3}\lambda_{a}\sigma^{\mu\nu}t_{R})\tilde{\phi}G^{a}_{\mu\nu},
 \end{eqnarray}
where $\tilde{\phi} = i\tau_{2}\phi^{*}$ and $\phi$ is the weak doublet of Higgs boson field, $q_{L3}$
is the quark weak doublet of left-handed quark and the right-handed top quark field is denoted 
by $t_{R}$.
It is notable that no corrections from dimension-six operators are received by the $\gamma_{\mu}$ term in the
Eq.\ref{e1}.

They are connected to the
effective dimension-six operator couplings through the following relations \cite{r2}:
\begin{eqnarray}\label{cgtt}
\delta d^{g}_{V} = \frac{\sqrt{2}}{g_{s}} {\rm Re}C^{33}_{uG\phi}\frac{vm_{t}}{\Lambda^{2}}~,~\delta d^{g}_{A} = \frac{\sqrt{2}}{g_{s}}{\rm Im}C^{33}_{uG\phi}\frac{vm_{t}}{\Lambda^{2}},
\end{eqnarray}
where $v$ is the vacuum expectation value and is equal to 246 GeV.
The chromoelectric dipole moment $d^{g}_{A}$ is corresponding to the
imaginary part of $C^{33}_{uG\phi}$. In this study, we consider both chromoelectric and chromomagnetic dipole moments.

In the SM context, the one-loop level  QCD corrections can generate
 $d^{g}_{V}$ through the exchange of gluons in two different Feynman
diagrams.  One of the diagrams is the same as QED case, replacing photon
by gluon.  Another diagram consists of an external gluon interacting with the internal gluons
coming from the non-abelian nature of QCD.  The same as QED case, these diagrams generate
non-zero $d^{g}_{V}$ which is proportional to $\alpha_{s}/\pi$ \cite{r61}. 
It is worth indicating that in addition to QCD corrections, $Z$  and Higgs bosons
 exchange also  generate $d^{g}_{V}$. Including all SM contributions at one-loop, the value of $d^{g}_{V}$ 
 is equal to  $-7\times 10^{-2}$ and non-zero value for $d^{g}_{A}$  arises 
 from contributions from beyond one-loop and is quite small \cite{r61,r62}.

At present, there are both {\it direct} and {\it indirect} bounds on
the  chromomagnetic and chromoelectric dipole moments of the top quark.
The bound could be obtained from the inclusive and differential top quark pair
cross section measurements at the LHC and Tevatron.
In Ref.\cite{r18}, we have shown that in particular the presence of
top quark chromoelectric dipole moment increases the gluon-gluon fusion process contribution in $t\bar{t}$
production at the Tevatron and LHC. Bounds are derived on both top quark chromoelectric and 
chromomagnetic dipole moments using the measured ratio $\sigma(gg\rightarrow t\bar{t})/\sigma(pp\rightarrow t\bar{t})$
and $t\bar{t}$ mass spectrum at the Tevatron \cite{r18}.

The top pair events produced at the large invariant masses in proton-proton collisions at the center-of-mass energies
of 13, 14, and 100 TeV in the semi-leptonic channel have been studied to probe the top quark dipole moments in Ref.\cite{rsavi}.
It has been shown that in the boosted regime the QCD background can be considerably suppressed and stringent bounds
are achievable. The CMS collaboration has derived limits on these dipole moments
from the measured top pair spin correlation at the LHC at 8 TeV \cite{rcms}.

The single top quark production in association with a $W$ boson  ($tW$-channel) is shown to be
also a sensitive process to the top quark dipole moments \cite{r9,r17}.   Constraints have been 
obtained using the measured  cross section
of $tW$-channel at the LHC with the center-of-mass energy of 7 TeV using an integrated luminosity of 4.9 fb$^{-1}$.
 
Amongst all searches, the strongest limits on  $d_{V}^{g}$ and
$d_{A}^{g}$ come from  low energy probes like the neutron electric dipole moment ($d_{n}$) \cite{nedm}
and the rare decays of $B$  mesons \cite{r61}. The constraint on the top quark chromoelectric dipole moment
from $d_{n}$ is found to be: $|d_{A}^{g}| \leq 0.95\times 10^{-3}$ at $90\%$ confidence level (CL)  \cite{nedm}.
The measured branching fraction of $b\rightarrow s\gamma$ leads to the limits of 
$3.8\times 10^{-3} \leq d_{V}^{g} \leq 1.2 \times 10^{-3}$ at the $95\%$ CL \cite{r61}.

\subsection{$Zt\bar{t}$ vertex}

The effective   $Zt\bar{t}$  vertex considering the SM contributions and the ones come from 
dimension six operators can be written as \cite{r2}:
\begin{eqnarray}
\label{e2}
\mathcal{L}_{Zt\bar{t}} = &-&\frac{g}{2c_{W}}\bar{t}\gamma_{\mu}(X_{L}P_{L}+X_{R}P_{R}-2s^{2}_{W}Q_{t})tZ^{\mu} \nonumber \\
&-&\frac{g}{2c_{W}}\bar{t}\frac{i\sigma_{\mu\nu}q^{\nu}}{m_{Z}}(d^{Z}_{V}+id^{Z}_{A}\gamma_{5})tZ_{\mu},
\end{eqnarray}
where $m_{Z}$ and $Q_{t}$ are the $Z$ boson mass and the top quark electric charge, respectively.
In the SM at tree level, $X_{L} = 1$, $X_{R} = 0$, and $d_{V}^{Z} = d_{A}^{Z} = 0$.  The contributions to 
these $Zt\bar{t}$ coupling from the dimension six operators are:
\begin{eqnarray}
\delta d^{Z}_{A} = \sqrt{2}\times\text{Im}[c_{W}C^{33}_{uW}-s_{W}C^{33}_{uB\phi}]\frac{v^{2}}{\Lambda^{2}}, ~
\delta d^{Z}_{V} = \sqrt{2}\times\text{Re}[c_{W}C^{33}_{uW}-s_{W}C^{33}_{uB\phi}]\frac{v^{2}}{\Lambda^{2}}.
\end{eqnarray}
The contributions of dimension six operators to $X_{L}$ and $X_{R}$ are neglected in this analysis \cite{r2}.
The constraints on $d^{Z}_{A}$ and $d^{Z}_{V}$  could be translated into limits on the
combination of the effective operators.
The couplings $d^{Z}_{A}$ and $d^{Z}_{V}$ are the weak 
electric and magnetic dipole moments. The weak electric dipole moment
coupling is a CP violating coupling which appears at three-loops in the SM and
the coupling  $d^{Z}_{V}$ corresponds to the weak magnetic
dipole moment  and  is at the order of 
$10^{-4}$ in the SM framework \cite{r63,r64,r65,r66}.

There are  studies on $d^{Z}_{A}$ and $d^{Z}_{V}$ at the 
electron-positron colliders and at the LHC \cite{r25,r67} to constrain these couplings. 
In Ref.\cite{r67}, it has been shown that by combining the LEP1 
data at Z-pole with top pair cross section
measurements and the electroweak precision data, the degeneracy between the 
involving operators in $d^{Z}_{A}$ and $d^{Z}_{V}$ could be broken.

The top quark weak electric and magnetic dipole moments have
been investigated at the LHC and the ILC from the $t\bar{t}Z$ production \cite{r25}.
 Both weak dipole moments are expected to be 
constrained to $\pm 0.15$ using 300 fb$^{-1}$ of data and would be improved
 to $\pm0.08$ with 3 ab$^{-1}$ integrated luminosity of the data. 
 Bounds at the same order can be obtained using the LEP electroweak precision data.
 The ILC with 500 fb$^{-1}$ is expected to reach the limits of $\pm 0.08$ on the weak electric dipole moment  and $[-0.02,0.04]$ on the
 weak magnetic dipole moment \cite{r25}.
It has been shown in  Ref.\cite{r23} that $d^{Z}_{V}$ and $d^{Z}_{A}$ can be well probed by the ratio of the cross section of
$t\bar{t}Z$ to $t\bar{t}$, because it allows to reduce several sources of the systematic uncertainties considerably.

\section{LHC constraints from $t\bar{t}VV$} \label{sns}

The $pp\rightarrow t\bar{t}WW$ and 
$pp\rightarrow t\bar{t}ZZ$ processes are interesting to study because of their small production cross sections in the SM \cite{r56}
and the significant enhancement that could show up in their rates in several new physics scenarios. 
In this section, we examine the sensitivity of these processes 
to the strong and weak top quark 
electric and magnetic dipole moments at the 14 TeV LHC.

The SM $t\bar{t}WW$ and $t\bar{t}ZZ$ processes produce an interesting set of the final states 
from which most of them giving rise to important signatures at the LHC.
For $t\bar{t}WW$ ($t\bar{t}ZZ$) process, depending on the top quarks and $W$ ($Z$) bosons decays 
between zero to four (six) charged lepton(s) might be produced.
Lists of  $t\bar{t}WW$ and $t\bar{t}ZZ$ decay modes, with at least a charged lepton in the final state,
and the related
branching fractions are presented in Table \ref{br_ttww} and Table \ref{br_ttzz}, respectively.

\begin{table}[]
\centering
\caption{$t\bar{t}WW$  decay modes where at least a charged lepton in the final state is present.}
\label{br_ttww}
\begin{tabular}{c|c|c|c} \hline
\textbf{$t\bar{t}$ decays}                      & \textbf{$WW$ decays}    & \textbf{Channel} & \textbf{Branching fraction}$\%$ \\  \hline
$(l\nu b)(q\bar{q}'b)$                            & $(q\bar{q}')(q\bar{q}')$ & \textit{mono-lepton}      &      20                                           \\
$(l\nu b)(q\bar{q}'b)$                            & $(l\nu)(q\bar{q}')$        & \textit{dilepton(OS+SS)}  &      20                                           \\
$(l\nu b)(q\bar{q}'b)$                            & $(l\nu)(l\nu)$               & \textit{trilepton}        &               4.8                                   \\ \hline
$(l\nu b)(l\nu b)$                                 & $(q\bar{q}')(q\bar{q}')$ & \textit{dilepton(OS)}     &           4.8                                      \\
$(l\nu b)(l\nu b)$                                 & $(l\nu)(q\bar{q}')$        & \textit{trilepton}        &                4.8                                 \\
$(l\nu b)(l\nu b)$                                 & $(l\nu)(l\nu)$              & \textit{four-lepton}     &                 1.2                               \\  \hline
$(q\bar{q}'b)(q\bar{q}'b)$                    & $(l\nu)(l\nu)$                & \textit{dilepton(OS)}                          &      4.8          \\
$(q\bar{q}'b)(q\bar{q}'b)$                    & $(l\nu)(q\bar{q}')$                & \textit{mono-lepton}                    &       20                              \\  \hline

\end{tabular}
\end{table}

For the $t\bar{t}WW$ process, the main decay channel is the mono-leptonic decay mode
which has a branching fraction of $40\%$, followed by the dilepton, opposite-sign and same-sign (OS+SS) mode with branching fraction of
$29.6\%$.
The branching fractions of trilepton and four-lepton decay modes are $9.6\%$ and $1.2\%$, respectively.
Among all the above decay modes the mono-leptonic suffers from large background contributions. 
The  channels in particular containing at least a pair of SS charged leptons  seem to be the 
promising search channels for the  $t\bar{t}WW$ process. 
For the $t\bar{t}ZZ$ process, in addition to mono-lepton, dilepton, trilepton, and four-lepton channels
five and six lepton multiplicities are among the possible decay channels. Although the topologies with 
high lepton multiplicities have small branching fractions, the contributing backgrounds for such cases are quite negligible.

In order to study the sensitivity of the $t\bar{t}WW$ and $t\bar{t}ZZ$ processes  to the
top quark strong and weak dipole moments, we employ \texttt{MadGraph5\_aMC@NLO} package \cite{mg5}
which  automatically generates the necessary code for computing the cross section and 
other observables for the related process.
The results are computed using the NNPDF3 PDF sets \cite{nnpdf}.
The top quark mass is set to 172.5 GeV and 
the mass of $W$ boson is taken as 80.37 GeV. 
The calculations are performed at the LHC with the center-of-mass energy of 14 TeV.

\begin{table}[]
\centering
\caption{$t\bar{t}ZZ$  decay modes where at least a charged lepton in the final state is present.}
\label{br_ttzz}
\begin{tabular}{c|c|c|c} \hline
\textbf{$t\bar{t}$ decays}                      & \textbf{$ZZ$ decays}    & \textbf{Channel} & \textbf{Branching fraction} \\  \hline
$(l\nu b)(q\bar{q}'b)$                            & $(q\bar{q})(q\bar{q})$ & \textit{mono-lepton}      &           21                                    \\
$(l\nu b)(q\bar{q}'b)$                            & $(\nu\bar{\nu})(\nu\bar{\nu})$ & \textit{mono-lepton}      &         1.76                                       \\
$(l\nu b)(q\bar{q}'b)$                            & $(l^{+}l^{-})(q\bar{q})$        & \textit{trilepton}  &                          6                       \\
$(l\nu b)(q\bar{q}'b)$                            & $(l^{+}l^{-})(l^{+}l^{-})$               & \textit{five-lepton}        &        0.43                                         \\  
$(l\nu b)(q\bar{q}'b)$                            & $(l^{+}l^{-})(\nu\bar{\nu})$               & \textit{trilepton}        &        1.75                                         \\  
$(l\nu b)(q\bar{q}'b)$                            & $(q\bar{q})(\nu\bar{\nu})$               & \textit{mono-lepton}        &        12.2                                         \\  \hline
$(l\nu b)(l\nu b)$                                 & $(q\bar{q})(q\bar{q})$ &        \textit{dilepton(OS)}     &                5.18                                 \\
$(l\nu b)(l\nu b)$                                 & $(\nu\bar{\nu})(\nu\bar{\nu})$ &        \textit{dilepton(OS)}     &       0.43                                        \\
$(l\nu b)(l\nu b)$                                 & $(l^{+}l^{-})(q\bar{q})$        & \textit{four-lepton}        &               1.48                                   \\
$(l\nu b)(l\nu b)$                                 & $(l^{+}l^{-})(l^{+}l^{-})$              & \textit{six-lepton}     &                  0.1                              \\  
$(l\nu b)(l\nu b)$                                 & $(q\bar{q})(\nu\bar{\nu})$              & \textit{dilepton}     &                  3                              \\  
$(l\nu b)(l\nu b)$                                 & $(l^{+}l^{-})(\nu\bar{\nu})$              & \textit{four-lepton}     &                  0.43                              \\  \hline
$(q\bar{q}'b)(q\bar{q}'b)$                    & $(l^{+}l^{-})(q\bar{q}')$        & \textit{dilepton(OS)}                          &     6.1           \\
$(q\bar{q}'b)(q\bar{q}'b)$                    & $(l^{+}l^{-})(\nu\bar{\nu})$              & \textit{dilepton(OS)}                    &       1.7                                \\  
$(q\bar{q}'b)(q\bar{q}'b)$                    & $(l^{+}l^{-})(l^{+}l^{-})$              & \textit{four-lepton}                    &       0.44                               \\  \hline
\end{tabular}
\end{table}

To perform the calculations of the cross sections in the presence of the 
top quark strong and weak dipole moments, the
effective Lagrangians are implemented
into \texttt{FeynRules} program \cite{feyn}. Then the effective model is exported 
into a UFO module \cite{ufo} which is connected to \texttt{MadGraph5\_aMC@NLO}.
\texttt{MadSpin} is used to for decaying top quarks, $W$ and $Z$ bosons.
\texttt{Pythia 8} \cite{pythia8} is used for parton showering and hadronization.
Jets are reconstructed using the anti-$k_{t}$ algorithm with a radius size of 0.4 \cite{jet}.
 All the results presented in this study are idealized and no object reconstruction and detector effects
are included. These effects modify the shape of final state distributions, however the study
of such effects is beyond the scope of this exploratory work and are left to a future analysis.

\subsection{Top pair production in association with two charged gauge bosons $W^{\pm}W^{\mp}$} \label{tw}

\begin{figure}[htb]
\begin{center}
\vspace{1cm}
\resizebox{0.45\textwidth}{!}{\includegraphics{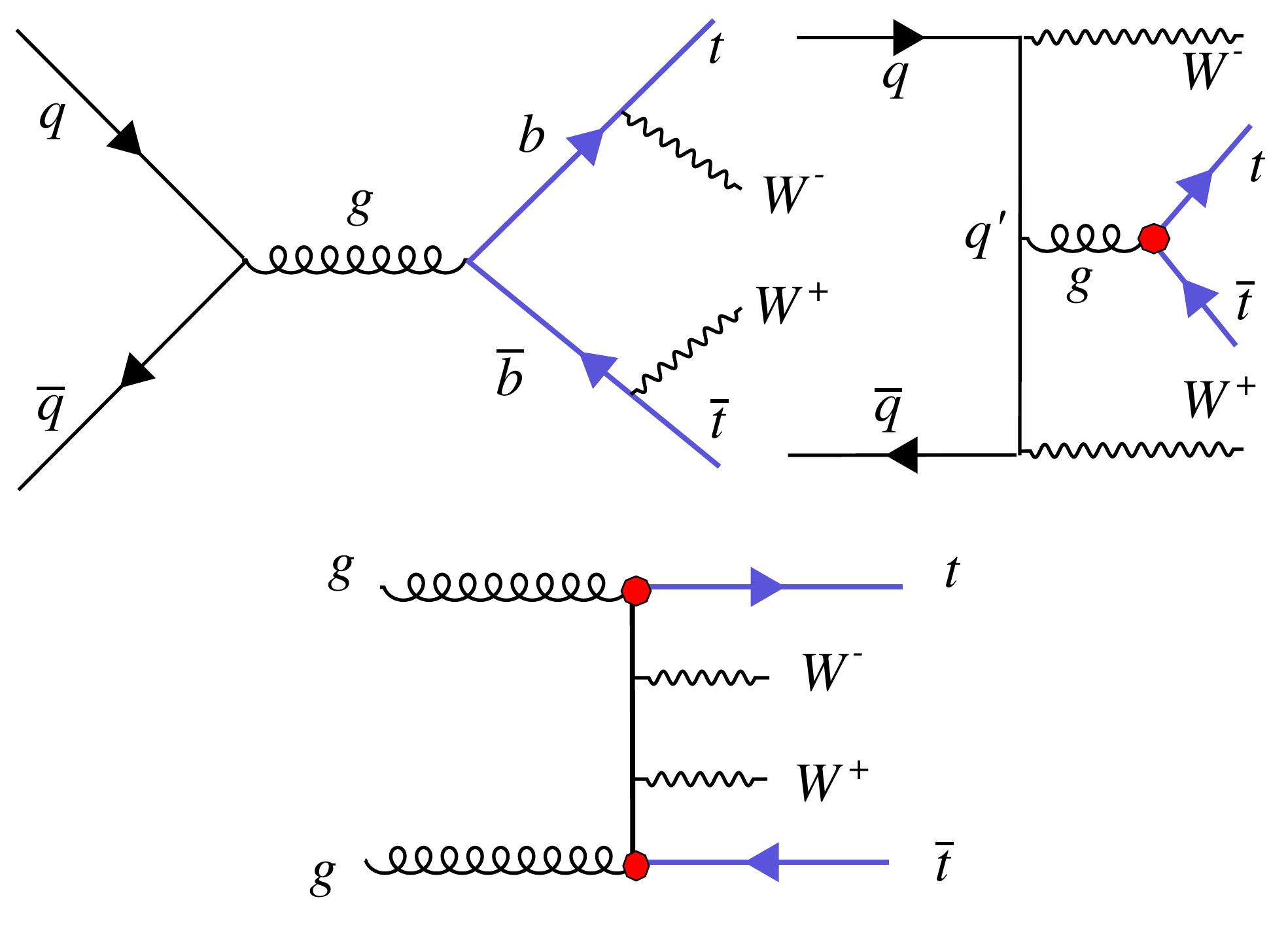}}  
\caption{  Lowers-order representative Feynman diagrams for $t\bar{t}WW$ production at the LHC. The vertices which receive 
contribution from $O^{33}_{uG\phi}$ operator are shown with red filled circles. }\label{feynmanww}
\end{center}
\end{figure}

In the SM, the production of  top quark pair associated with $W^{\pm}W^{\mp}$ come from either 
gluon-gluon fusion or quark-anti-quark annihilation.  The main contributions are of order $\mathcal{O}(\alpha_{s}^{2}\alpha^{2})$
and a partonic center-of-mass energy of at least $2m_{t}+2m_{W}$ is necessary which causes
a  small production cross section at the LHC.  Gluon and quark initiated representative Feynman diagrams at leading order
contributing to $t\bar{t}WW$ production in the SM are depicted in Fig.~\ref{feynmanww}.
In our study the production of $t\bar{t}WW$ is calculated in four flavor scheme (4FS)
as in the 5FS case there exists intermediate top quark resonances that must be subtracted \cite{r56,r80}. It is to avoid of 
unnecessary complication in calculation of the production rate.

The $t\bar{t}WW$
 production cross section at the center-of-mass energy of 14 TeV
is calculated using \texttt{MadGraph5\_aMC@NLO} package. The next-to-leading order cross section 
 is found to be $14.5~{\text {fb}}\pm 3\%$ (PDF)$^{+12.3\%}_{-13.0\%}$ (scales). The  
NLO QCD effects are on the order of $10\%$.  Complete details of  the QCD NLO calculations can be found in Refs.\cite{r80}.
We note that at the 14 TeV LHC, around $54\%$ of the total cross section comes from the
gluon-gluon fusion which goes higher at the larger center-of-mass energies because of 
 growing of the gluon PDF.

The LO contributions of the top quark chromoelectric ($d_{A}^{g}$) and chromomagnetic ($d_{V}^{g}$) dipole moments, arising from
$O^{33}_{uG\phi}$ operator, 
to the $t\bar{t}WW$ rate is calculated with  \texttt{MadGraph5\_aMC@NLO}. 
The relative corrections 
from $d_{A}^{g}$ and $d_{V}^{g}$ to the total cross section of $\sigma(pp\rightarrow t\bar{t}WW)$ has 
the following form:
\begin{eqnarray}\label{cs}
\frac{\Delta\sigma(pp\rightarrow t\bar{t}WW)}{\sigma_{\rm SM}} = \alpha_{i} d_{i}^{g} + \beta_{i} (d_{i}^{g} )^{2}~,~ i = V,A,
\end{eqnarray}
where $\sigma_{SM}$ is the SM cross section and 
$\alpha_{i}$  is the interference term which its contribution is of the order of $\Lambda^{-2}$.
The $\beta_{i}$ term corresponds to the pure $O^{33}_{uG\phi}$ contributions
which has the power of $\Lambda^{-4}$. 
Without taking into account the dimension eight operators, such terms could be dropped
because dimension eight operators generate contributions at similar order.
However, we keep $\Lambda^{-4}$ term as it is the first appearing term in the cross section for $d_{A}^{g}$ and it is relevant to
have it when obtaining constraints on $d_{V}^{g}$.
Of course, it is expected that the cross section has a symmetric shape around $d_{A}^{g} = 0$ as it is a CP even observable
leading to $\alpha_{A} = 0$.
To extract the coefficients $\alpha_{i}$ and $\beta_{i}$ in Eq.\ref{cs}, the calculations with $d_{A}^{g}$ and $d_{V}^{g}$
are performed assuming different values: $0.0, \pm 01,\pm 0.2, \pm 0.3$ and fit the obtained 
cross sections to Eq.\ref{cs}. The coefficients $\alpha_{i}$ and $\beta_{i}$ 
are presented in Table \ref{xsecWW}.

\begin{table}[]
\centering
\caption{Values of $\alpha_{i}$ and $\beta_{i}$  for the 14 TeV LHC.}
\label{xsecWW}
\begin{tabular}{lll} \hline\hline
$i$ & $\alpha_{i}$ & $\beta_{i}$ \\ \hline\hline
$V$ &        -1.2      &     1783.5        \\
$A$ &       0.0       &      1950.8      \\ \hline
\end{tabular}
\end{table}

To derive a quantitative estimate of the constraints that could be optimistically reached under various
integrated luminosity scenarios, we concentrate on the exactly two same sign charged lepton ($e,\mu$) topology. 
To select the same sign dilepton events, we require to have exactly two SS leptons with transverse momentum greater 
than 20 GeV and $|\eta_{l}| < 2.5$.  The angular separation of the leptons, 
$\Delta R(l_{1},l_{2}) = \sqrt{(\Delta \eta)^2+(\Delta \phi)^{2}}$, is requested to larger than 0.4. 
The event is required to contain at least four jets from which at least two have to be matched with a B-hadron. 
We  continue  to  set  an  upper  limit  on  the  $t\bar{t}WW$  production
cross  section  in  the  presence  of strong chromoelectric or chromomagnetic dipole moments.

To derive constraints on $d_{V}^{g}$ and $d_{A}^{g}$, a counting experiment technique
is employed. The method is to begin with a Poisson distribution describing the probability for measuring $N$ events:
\begin{eqnarray}\label{l1}
\mathcal{P}(N|\sigma_{t\bar{t}WW}\times \epsilon \times \mathcal{L},B) = e^{-(\sigma_{t\bar{t}WW}\times \epsilon \times \mathcal{L}+B)}\times \frac{(\sigma_{t\bar{t}WW}\times \epsilon \times \mathcal{L}+B)^{N}}{N!},
\end{eqnarray}
where $\sigma_{t\bar{t}WW}$, $\mathcal{L}$, $\epsilon$ and $B$
are the signal cross section in the presence of $d_{V}^{g}$ and $d_{A}^{g}$,
the integrated luminosity, the efficiency of signal after the selection criteria, and the expected background events corresponding to 
the assumed integrated luminosity.
At $ 95\%$ confidence level (CL), the upper limit on the signal cross section can be calculated with integration 
over the posterior probability according to the following:
\begin{eqnarray}\label{l2}
0.95 = \frac{\int_{0}^{\sigma^{95\%} }\mathcal{P}(N|\sigma_{t\bar{t}WW}\times \epsilon \times \mathcal{L},B)}
{\int_{0}^{\infty }\mathcal{P}(N|\sigma_{t\bar{t}WW}\times \epsilon \times \mathcal{L},B)}.
\end{eqnarray}
In this exploratory study, the number of background events is obtained as $B = (\sigma^{\rm SM}_{t\bar{t}WW} + \sigma^{\rm SM}_{t\bar{t}W})\times \mathcal{L}$ 
where $\sigma^{\rm SM}_{t\bar{t}WW}$  and $\sigma^{\rm SM}_{t\bar{t}W}$ are the SM production rate for $t\bar{t}WW$ and  $t\bar{t}W$ processes after the selection cuts described above. 
To be more realistic, the SM production cross section of these backgrounds are scaled to their NLO value.
Assuming $60\%$ b-tagging efficiency and full efficiency for lepton reconstruction, the efficiency $\epsilon$ is found to be $33\%$.
To have a realistic estimation of the efficiency $\epsilon$, a detailed experimental simulation to consider full detector response
must be done which is beyond the scope of this study. 

We obtain the expected upper limit at the $95\%$ CL on the signal cross section and compare it with the theoretical signal cross section
to find the upper limits on $d_{V}^{g}$ and $d_{A}^{g}$. The resulting limits are calculated for three scenarios of integrated luminosities of
30, 300, 3000 fb$^{-1}$ and presented in Table \ref{dfdf}.
For example, with an integrated luminosity of 30 fb$^{-1}$ the upper limits of $ -0.027 \leq d_{V}^{g}  \leq 0.028$ and $|d_{A}^{g}| \leq 0.026$
are derived.
If we assume $10\%$ uncertainty on the signal efficiency and $100\%$ uncertainty on the
number of background events, the bounds on $d^{g}_{V}$ and $d^{g}_{A}$ at 30 fb$^{-1}$ are loosen to $-0.037  \leq d^{g}_{V} \leq 0.038$
and $  |d^{g}_{A}| \leq 0.036$.

\begin{table}[]
\centering
\caption{Limits on $d^{g}_{V}$ and $d^{g}_{A}$ at $95\%$ CL  corresponding to 30, 300, and 3000 fb$^{-1}$ integrated luminosities.}
\label{dfdf}
\begin{tabular}{llll} \hline\hline
Coupling       & 30 fb$^{-1}$                                   &  300 fb$^{-1}$ &  3000 fb$^{-1}$  \\ \hline\hline
$d^{g}_{V}$ &        [-0.026,0.027]   &     [-0.014,0.015]     &   [-0.008,0.009]   \\
$d^{g}_{A}$ &       [-0.025,0.025]       &      [-0.014,0.014]    &    [-0.008,0.008]   \\ \hline
\end{tabular}
\end{table}

We note that including the other signatures of $t\bar{t}WW$ process such as trilepton and four lepton would 
increase the sensitivity of this channel to the strong electric and magnetic dipole moments of the top quark.
In the end of this section,  it should be indicated that in addition to $gt\bar{t}$ effective couplings,
$t\bar{t}WW$ process is sensitive to the anomalous $Wtb$ and $Zt\bar{t}$ vertices.
The effective Lagrangian up to dimension six operators explaining the anomalous $Wtb$ coupling as follows \cite{r2}:
\begin{eqnarray}
\mathcal{L}_{\rm Wtb} = -\frac{g}{\sqrt{2}}\bar{b}\bigg(\gamma^{\mu}(V_{L}P_{L}+V_{R}P_{R})+
\frac{i\sigma_{\mu\nu}q^{\nu}}{m_{W}}(g_{L}P_{L}+g_{R}P_{R})\bigg)tW_{\mu}^{-}+h.c.,
\end{eqnarray}
where $V_{L,R}$ and  $g_{L,R}$ are dimensionless couplings. At tree level within the SM, 
$V_{L} = V_{tb}$ and $V_{R} = g_{L}=g_{R} = 0$.
From the rare B-meson decay, the constraints on these couplings are found to be \cite{bbb}:
 \begin{eqnarray}
 -0.0007< V_{R} < 0.0025~,~ -0.0013 < g_{L} < 0.0004, ~  -0.15 <   g_{R}  < 0.57.
 \end{eqnarray}
 The $95\%$ CL bounds derived from $W$ boson polarization and measured cross section of 
 the single top t-channel at the LHC are \cite{ccc}: $-0.13 < V_{R}  < 0.18$, $-0.09 < g_{L}   < 0.06$, and $-0.15 < g_{R} < 0.01$. 
The total cross section of $t\bar{t}WW$ process does not show considerable  sensitivity to $g_{L}$ and $g_{R}$.
By setting $g_{R} = 0.1$ and  $g_{L} = 0.1$, the relative change of  $t\bar{t}WW$ rate is $4\%$  and $0.24\%$, respectively. This
means that no strong limits on $g_{L}$ and $g_{R}$ are expected to be obtained from $t\bar{t}WW$ channel \footnote{The correct prediction for 
examining the sensitivity of the $t\bar{t}WW$ process to the anomalous $Wtb$ should be performed by 
including the top quark decays since two additional $Wtb$ vertices appear. }.

We also note that in $t\bar{t}WW$ production, there are diagrams 
containing $Zt\bar{t}$ vertex resulting to the fact that the weak dipole moments $d^{Z}_{V}$ and 
$d^{Z}_{A}$ contribute to $t\bar{t}WW$ cross section. We do not consider this in our analysis as the modification
to $\sigma(pp \rightarrow t\bar{t}WW)$ due to  $d^{Z}_{V,A}$
is found to be at the level of less than $10\%$ when these couplings vary up to the value of $\pm 0.05$.

\subsection{Top pair production in association with two neutral heavy gauge bosons $ZZ$} \label{tz}

\begin{figure}[htb]
\begin{center}
\vspace{1cm}
\resizebox{0.45\textwidth}{!}{\includegraphics{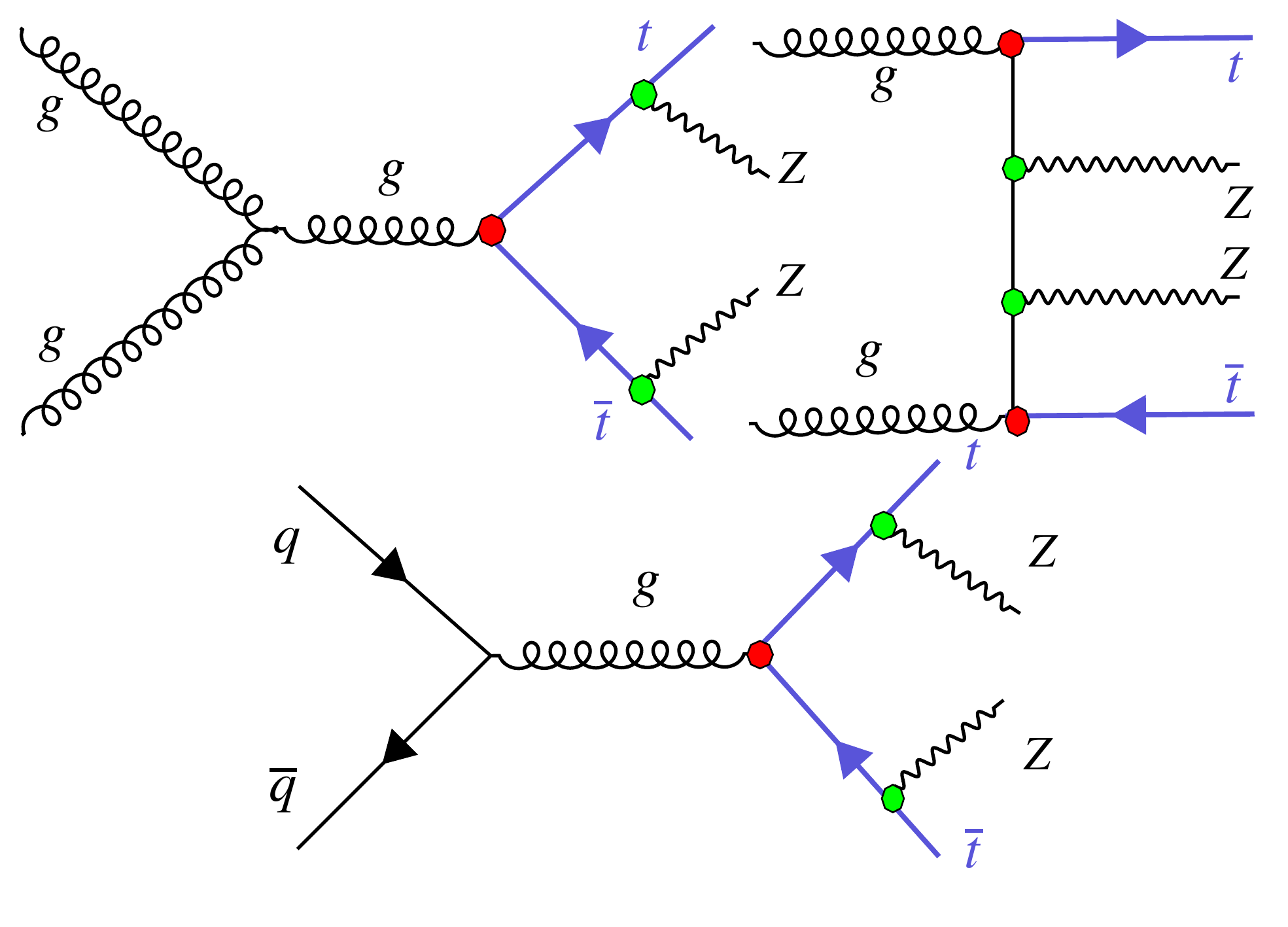}}  
\caption{  Representative Feynman diagrams for $t\bar{t}ZZ$ production at leading-order.  
}\label{feynmanzz}
\end{center}
\end{figure}

In this section, we study the sensitivity of the  $t\bar{t}ZZ$ production to the top quark dipole moments.
The representative Feynman diagrams at leading order of this process are depicted in Fig.\ref{feynmanzz}.
The next-to-leading-order cross section of $t\bar{t}ZZ$ process is calculated using
\texttt{MadGraph5\_aMC@NLO}  is found to be: $2.6~{\text {fb}}\pm 1.82\%$ (PDF)$^{+4.34\%}_{-8.78\%}$ (scales), where
the first uncertainty gives the contribution from the dependence on the choice of parton
distribution functions and the second part is the  factorization and renormalization scale uncertainties \cite{r56,r80}.
The input parameters for the cross section calculation has been taken similar to the previous section.
The NLO corrections to the $t\bar{t}ZZ$ production is quite small resulting to a $k$-factor 
close to one \cite{r80}. The leading order cross section is proportional to $\mathcal{O}(\alpha_{s}^{2}\alpha^{2})$ 
and a partonic center-of-mass energy of at least $2m_{t}+2m_{Z}$ is necessary for such a final state at the LHC.
The presence of $\alpha^{2}$ and four heavy particles in the final state, which causes to reduce the phase space,
lead to such a small rate for this process.

The $t\bar{t}ZZ$ channel allows us to probe both the strong ($d^{g}_{V,A}$) and weak ($d^{Z}_{V,A}$) top quark dipole moments.
  The contributions of the strong and weak dipole moments  
to the $t\bar{t}ZZ$ productions cross section is calculated using \texttt{MadGraph5\_aMC@NLO} package. 
The relative modifications from  operators $O^{33}_{uG\phi}$, $O^{33}_{uW}$ and $O^{33}_{uB\phi}$
to the total cross section of $\sigma(pp\rightarrow t\bar{t}ZZ)$ in terms of  $d^{g}_{V,A}$ and $d^{Z}_{V,A}$ can be written as:
\begin{eqnarray}\label{cszz}
\frac{\Delta\sigma(pp\rightarrow t\bar{t}ZZ)}{\sigma_{\rm SM}} = \rho^{g,Z}_{i} d_{i}^{g,Z} + \gamma^{g,Z}_{i} (d_{i}^{g,Z} )^{2}~,~ i = V,A,
\end{eqnarray}
where  $\rho^{g,Z}_{i}$ term is the interference term of the SM with new physics which is of the order of $\Lambda^{-2}$.
The $\gamma^{g,Z}_{i}$ term is corresponding to the pure $O^{33}_{uG\phi}$, $O^{33}_{uW}$ and $O^{33}_{uB\phi}$ contributions
appearing with the power of $\Lambda^{-4}$. 
To obtain the coefficients $\rho^{g,Z}_{i}$ and $\gamma^{g,Z}_{i}$ in Eq.\ref{cszz}, the 
cross sections are calculated in the presence of these coefficients
taking various values: $0.0, \pm 01,\pm 0.2, \pm 0.3$, then the results are fitted to Eq.\ref{cszz}. The coefficients  $\rho^{g,Z}_{i}$ and $\gamma^{g,Z}_{i}$
are given in Table \ref{xsecZZzz}. We see the interference term coefficient (for $i=V$) is small and 
the pure new physics coefficients are almost close to each other.
 As expected due to the presence of $q_{\mu}$ factor in the effective Lagrangian, 
 the  coefficients $\gamma^{g,Z}_{V,A}$ are very large.

\begin{table}[]
\centering
\caption{Values of $\rho^{g,Z}_{i}$ and $\gamma^{g,Z}_{i}$ for the 14 TeV LHC.}
\label{xsecZZzz}
\begin{tabular}{lllll} \hline\hline
$i$ & $\rho^{g}_{i}$ & $\gamma^{g}_{i}$ &  $\rho^{Z}_{i}$ & $\gamma^{Z}_{i}$ \\ \hline\hline
$V$ &        -6.0      &     2127.2     &   0.1 &  27.5    \\
$A$ &       0.0       &      2092.4     &    0.0  &   27.8 \\ \hline
\end{tabular}
\end{table}

As mentioned before, there are several signatures for $t\bar{t}ZZ$
that all contain at least two b-jets which come from the weak top quark decay.
Among all signatures, we take the four-lepton (lepton $=e,\mu$) final state which 
is a clean signature.
Requiring four leptons and two b-tagged jets in the final state should be enough to increase
 the signal-to-background ratio significantly. To select the signal events,
 we require to have exactly four leptons with $p_{T} > 10$ GeV and $|\eta_{}| < 2.5$. The missing 
 transverse energy has to be larger than 30 GeV and each event  is requested to contain at least two b-tagged jets.
 To have well isolated objects in the final state, it is required $\Delta R(l_{i},l_{j})  > 0.4$, $\Delta R(j_{i},j_{j})  > 0.4$,
 and $\Delta R(l_{i},j_{j})  > 0.4$.

We follow the same method as described in the previous section to set upper limit on the 
signal cross section then the upper limit is translated into the limits on the top 
quark dipole moments.  The SM $t\bar{t}ZZ$  and $t\bar{t}Z$ 
are taken as the main backgrounds and the number of background events is obtained through 
$B = (\sigma^{\rm SM}_{t\bar{t}ZZ}+\sigma^{\rm SM}_{t\bar{t}Z})  \times \mathcal{L}$ 
where $\sigma^{\rm SM}_{t\bar{t}ZZ}$ and $\sigma^{\rm SM}_{t\bar{t}Z}$ are the SM  rates after the selection cuts described above. 
Taking a $60\%$ b-tagging efficiency and fully efficient lepton reconstruction, the efficiency $\epsilon$ is obtained to be equal to $32\%$.
The bounds on $d^{g}_{V,A}$ and $d^{Z}_{V,A}$ are shown in Table \ref{zzbounds} for 
30, 300 and 3000 fb$^{-1}$ integrated luminosity of data.

\begin{table}[]
\centering
\caption{Limits on $d^{g,Z}_{V}$ and $d^{g,Z}_{A}$ at $95\%$ CL  corresponding to 30, 300, and 3000 fb$^{-1}$ integrated luminosities.}
\label{zzbounds}
\begin{tabular}{llll} \hline\hline
Coupling       & 30 fb$^{-1}$                                   &  300 fb$^{-1}$ &  3000 fb$^{-1}$  \\ \hline\hline
$d^{g}_{V}$ &        [-0.023,0.026]   &     [-0.012,0.015]     &   [-0.006,0.009]   \\
$d^{g}_{A}$ &       [-0.024,0.024]       &      [-0.013,0.013]    &    [-0.007,0.007]   \\ \hline
$d^{Z}_{V}$ &     [-0.22,0.21]         &   [-0.12,0.11]      &   [-0.07,0.06]  \\
$d^{Z}_{A}$ &     [-0.21,0.21]          &  [-0.11,0.11]       & [-0.06,0.06]  \\ \hline
\end{tabular}
\end{table}

Assuming a $10\%$ overall uncertainty on the efficiency of signal and $100\%$
uncertainty on the number of background events make limits looser. Using 30 fb$^{-1}$
integrated luminosity of data, the bounds on $d^{g}_{V}$ and $d^{g}_{A}$ become  
$ -0.033 \leq d^{g}_{V} \leq 0.036 $ and $ |d^{g}_{A}|  \leq 0.035$.

One can derive a lower limit on the new physics characteristic scale using the Eq.\ref{cgtt} and
taking the Wilson coefficient  $C^{33}_{uG\phi}$ to be at most
equal to $4\pi$. Using for instance the obtained upper limit on $d^{g}_{V}$ at 3000 fb$^{-1}$
, a lower bound of  $\Lambda \sim 9$ TeV is deduced.
 Of course, choosing lower value of $C^{33}_{uG\phi}$ leads to looser limit on $\Lambda$.

\subsection{Comparison of the results with other studies}

\begin{figure}[htb]
\begin{center}
\vspace{1cm}
\resizebox{0.45\textwidth}{!}{\includegraphics{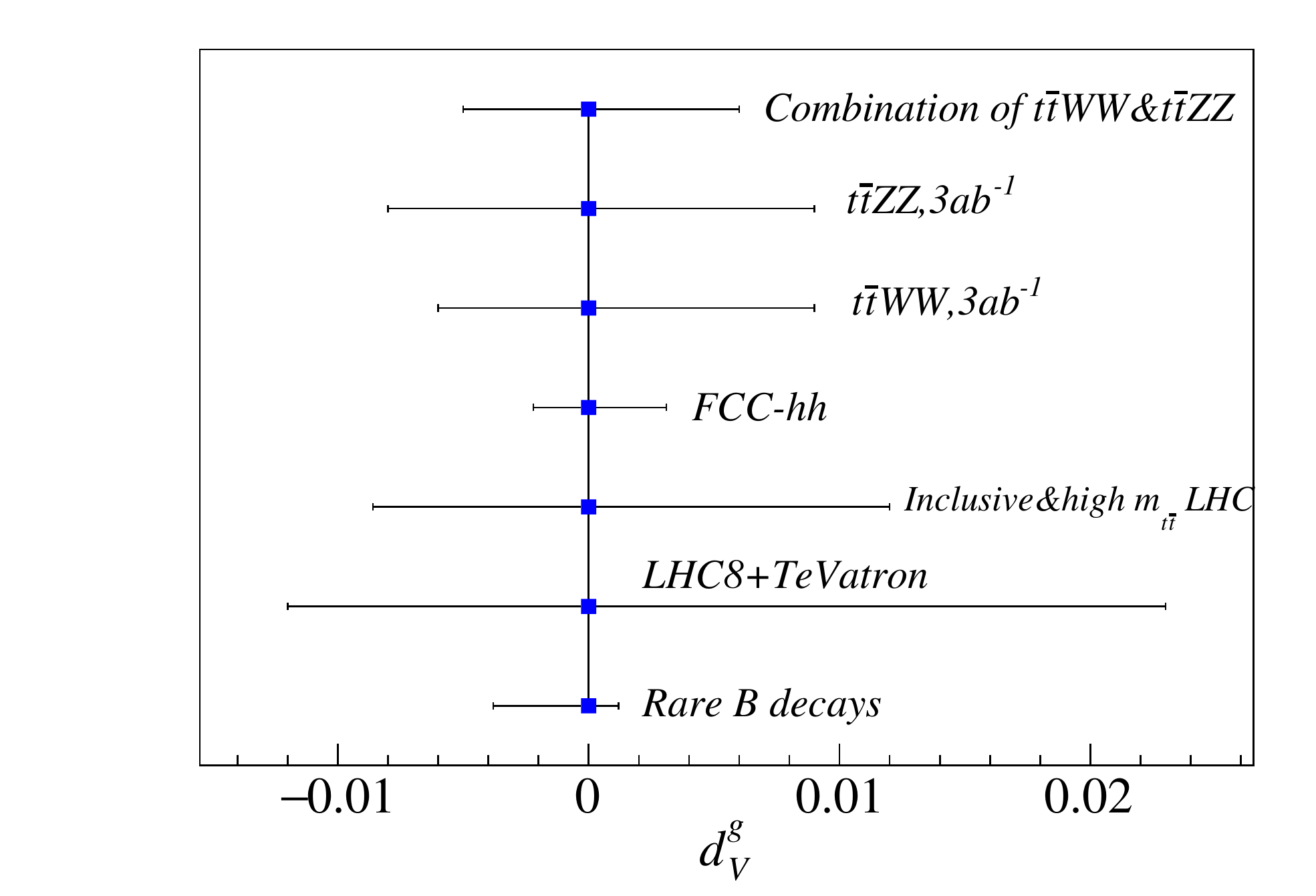}}  
\resizebox{0.42\textwidth}{!}{\includegraphics{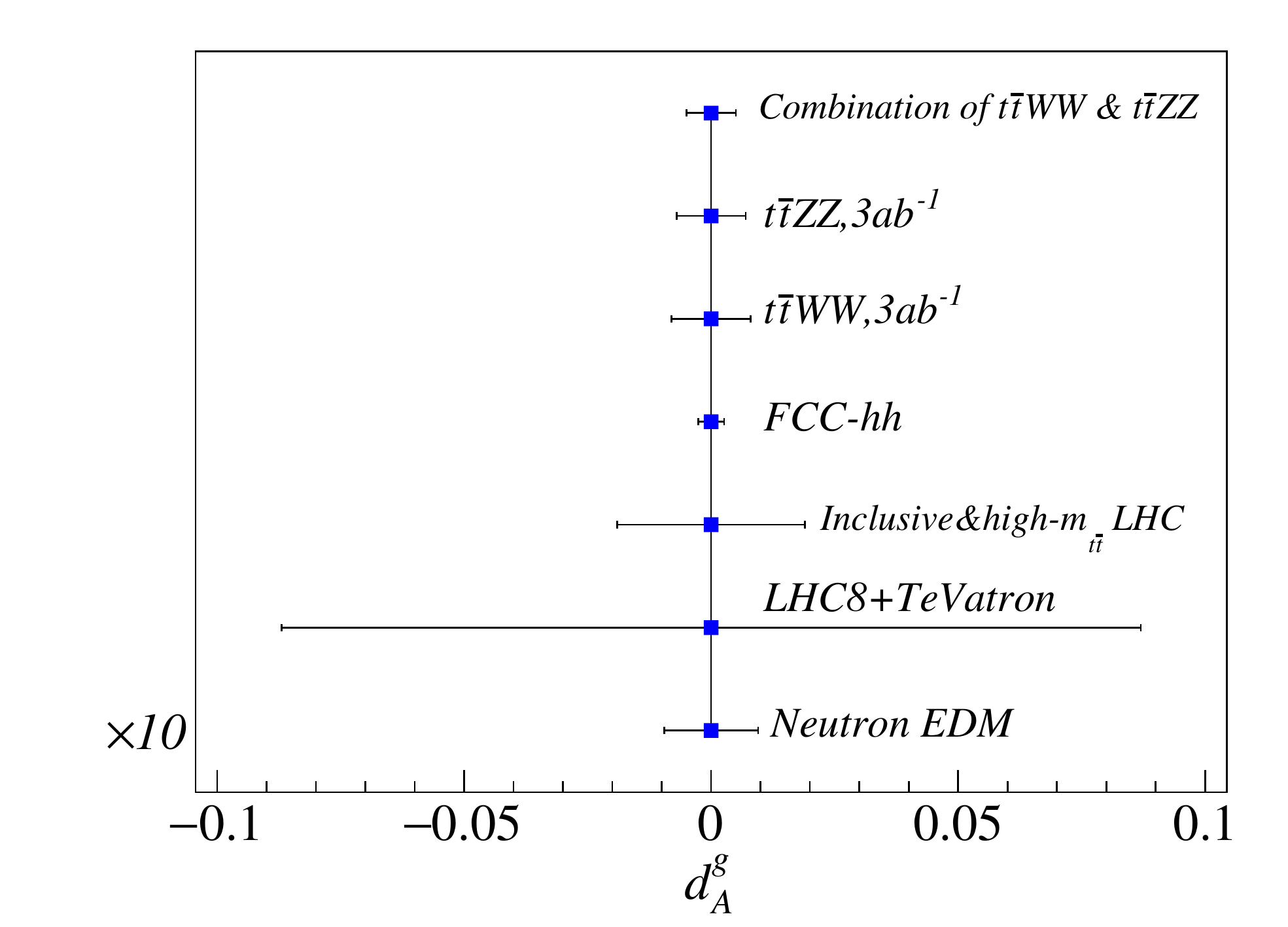}}  
\caption{ The limits at $95\%$ CL on  $d^{g}_{V}$ (right panel)  and on $d^{g}_{A}$ (left panel) from $t\bar{t}WW$ (same-sign leptons) and $t\bar{t}ZZ$ (four-lepton)
with 300 and 3000 fb$^{-1}$ are shown. The indirect limits on $d^{g}_{A}$ (neutron electric dipole moment) and on
$d^{g}_{V}$ (rare B meson decays) are presented as well as the limits from the combination of $t\bar{t}$ cross section at 
the LHC8 and Tevatron. Also, the limits which could be derived from tail of $t\bar{t}$ mass spectrum at the FCC-hh and LHC 
are shown.} \label{results1}
\end{center}
\end{figure}

In this section, we compare the sensitivity of the
expected constraints from the $t\bar{t}WW$ (same-sign leptons) analysis 
and $t\bar{t}ZZ$ (four-lepton) analysis with some other studies.
The results of this study  with two scenarios of 
integrated luminosities 300 and 3000 fb$^{-1}$ are compared with others
in Fig.\ref{results1}.  The most stringent direct bounds from the 
FCC-hh, where protons are collided with $\sqrt{s} = 100$ TeV, 
are based on the integrated luminosity of 10 ab$^{-1}$ \cite{rsavi} and are derived
from the events with central jets ($|\eta| < 2$) and transverse momentum 
larger than 1 TeV reconstructed using an anti-$k_{T}$ \cite{jet}
algorithm with a radius size of 0.2. The FCC-hh 
limits are obtained in an optimal invariant mass region of the
top quark pair mass of $m_{t\bar{t}} > 10$ TeV.

The indirect limits on $d^{g}_{V}$ are based on rare B meson decay \cite{r61}
which has been found to be $-0.0038 \leq d^{g}_{V} \leq 0.0012$. 
In particular, the upper limit is the most stringent one which is even 
stronger than the expected bound from FCC-hh. 
The combination of the measured top quark pair cross section
at the LHC8 and Tevatron lead to $-0.012 \leq d^{g}_{V} \leq 0.023$ \cite{rsavi}
and the expected limit derived from the $t\bar{t}$ spectrum and the inclusive 
cross section at the LHC14 based on 100 fb$^{-1}$ is $-0.0086 \leq d^{g}_{V} \leq 0.012$ \cite{rsavi}.
The limits from our analyses are comparable to these limits
and could be even improved if the other signatures presented in Table \ref{br_ttww} and Table \ref{br_ttzz}
are taken into account.

For the $d^{g}_{A}$ case, the indirect limits have been 
extracted from the upper limit on the neutron electric dipole moment. 
This indirect low energy limit which is  $|d^{g}_{A}| \leq 0.00095$  \cite{nedm} is the strongest one.
Again, among the direct limits, the one obtained from FCC-hh 
is the most stringent limit: $|d^{g}_{A}| \leq 0.0026$. 
The combination of the measured $t\bar{t}$ cross section
at the LHC8 and Tevatron implies  $ |d^{g}_{A}| \leq 0.087$ \cite{rsavi}
while the ones from $t\bar{t}$ spectrum and the inclusive 
cross section at LHC14 with 100 fb$^{-1}$ are $ |d^{g}_{A}| \leq 0.019$ \cite{rsavi}.

The combination of $t\bar{t}WW$ and  $t\bar{t}ZZ$ channels provides the limits of $-0.006 \leq d^{g}_{V} \leq 0.005$
and $ |d^{g}_{A}| \leq 0.005$ with an integrated luminosity of 3000 fb${-1}$.
The limits from $t\bar{t}WW$ (same-sign leptons), $t\bar{t}ZZ$ (four-lepton) and their combination
are comparable to the limits from other studies and in some cases would be even better.
The bounds obtained from this analysis could be improved by including the other signatures 
and taking into account the higher order QCD corrections in the signal channels.
It should be indicated that while the indirect limits
from the rare B decays and the neutron electric dipole moment
are stronger but they are complementing each other.

Now, we turn to the weak dipole moments $d^{Z}_{V}$ and $d^{Z}_{A}$.
The expected constraints from an electron-positron collider at $\sqrt{s} = 500$ GeV
with an integrated luminosity of 500 fb$^{-1}$,  are $|d^{Z}_{V}| \leq 0.08$ and 
$ -0.02 \leq d^{Z}_{A}| \leq 0.04$ \cite{r25}. These limits are derived by exploiting the total cross section 
of the top quark pair production. The limits from the LHC top pair production at the 
center-of-mass energy of 14 TeV with an integrated luminosity of 3 ab$^{-1}$
are $|d^{Z}_{V,A}| \leq 0.08$ \cite{r25}. They are obtained from the $p_{T, Z}$ distribution
in $t\bar{t}Z$ production. The expected limits  from the present study as shown 
in Table \ref{zzbounds} are comparable with the ones from ILC and LHC in $t\bar{t}Z$ channel.
At the end, it should be mentioned that 
our bounds are purely based on statistical sensitivity calculations and no experimental effects,
which would weaken them,  are taken into account. However, the combination of different decay channels for each process and
considering QCD higher order corrections would lead to have larger statistics and significant improvements.

\section{Sensitive observables}\label{ang}

Various types of vector, tensor and pseudo-tensor couplings in the effective Lagrangian of $gt\bar{t}$
and $Zt\bar{t}$ could lead to changes in differential distributions of the final state particles.
Therefore, one can exploit the differential rates to design 
analyses for achieving improvements with respect to the total cross sections.
Specially, it becomes very important for the cases that the reachable sensitivities from the 
total cross section in future prospects are not exciting enough.

There are already studies where the authors proposed 
several observables to reach more sensitivities to the 
effective couplings in the top quark sector and also to disentangle 
CP-even and CP-odd couplings \cite{r14,r39}.
Here, we construct an observable on the basis of 
the momenta of the final state in $t\bar{t}ZZ$ channel before 
decaying of the top quarks and the $Z$ bosons. 
Of course,  for the  reconstruction of  the $t\bar{t}ZZ$,  we need  full  information of the  top quarks and
$Z$ boson decay products momenta.
However, due to the presence of the missing neutrino when a top quark decays leptonically,
it is  impossible to reconstruct the top quark(s) completely and one needs to use $W$ boson 
and top mass as constraints to find the full momenta of the top quarks.
Furthermore,  ambiguities arise in combination of the decay products when assigning 
each particle to its real mother.
Such ambiguities would lead to large systematic uncertainties
to the $t\bar{t}ZZ$ system. In this exploratory work, we design 
observables using the momenta of $t\bar{t}ZZ$ and leave the explained
complications to a future study. 

\subsection{Invariant mass distributions}

\begin{figure}[htb]
\begin{center}
\vspace{1cm}
\resizebox{0.4\textwidth}{!}{\includegraphics{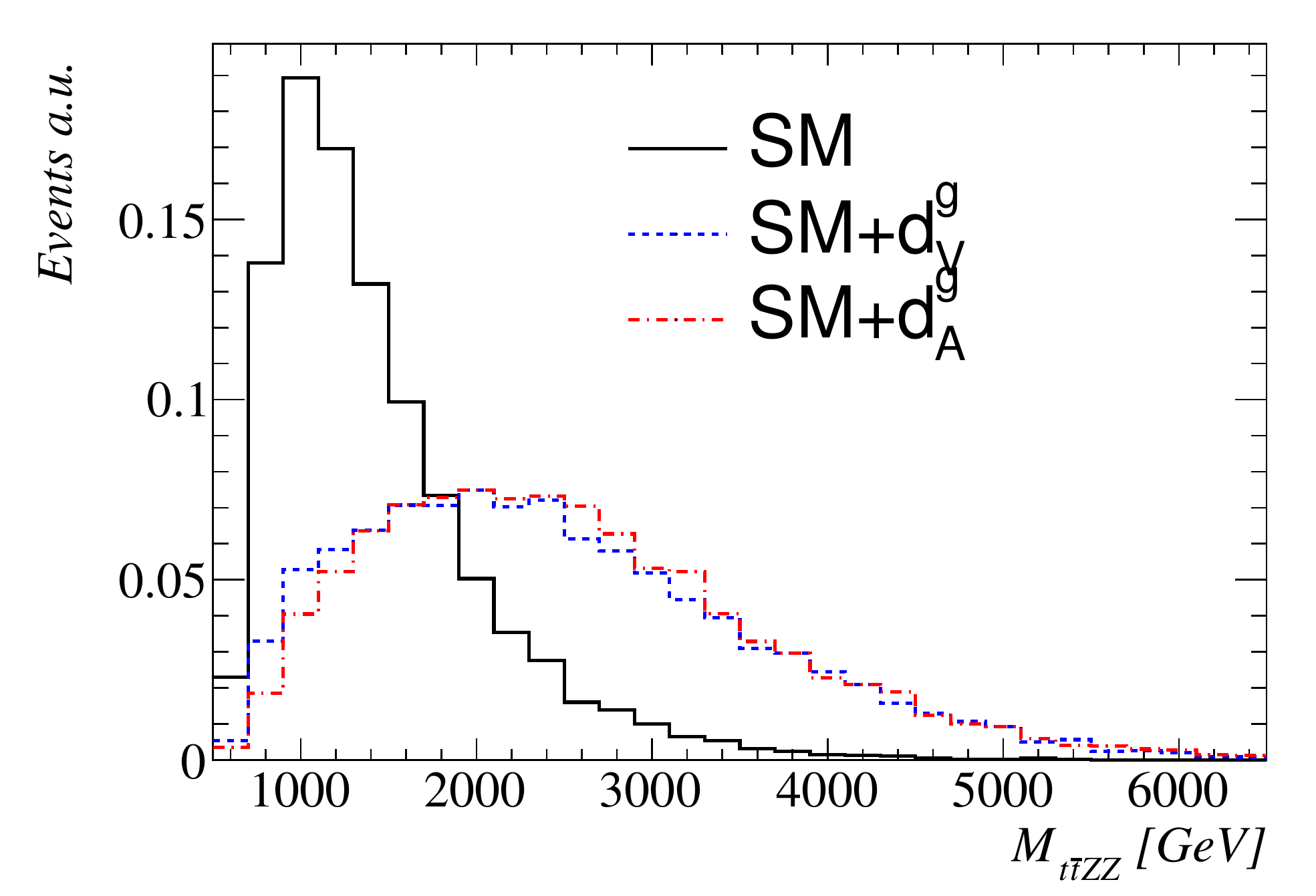}}  
\resizebox{0.4\textwidth}{!}{\includegraphics{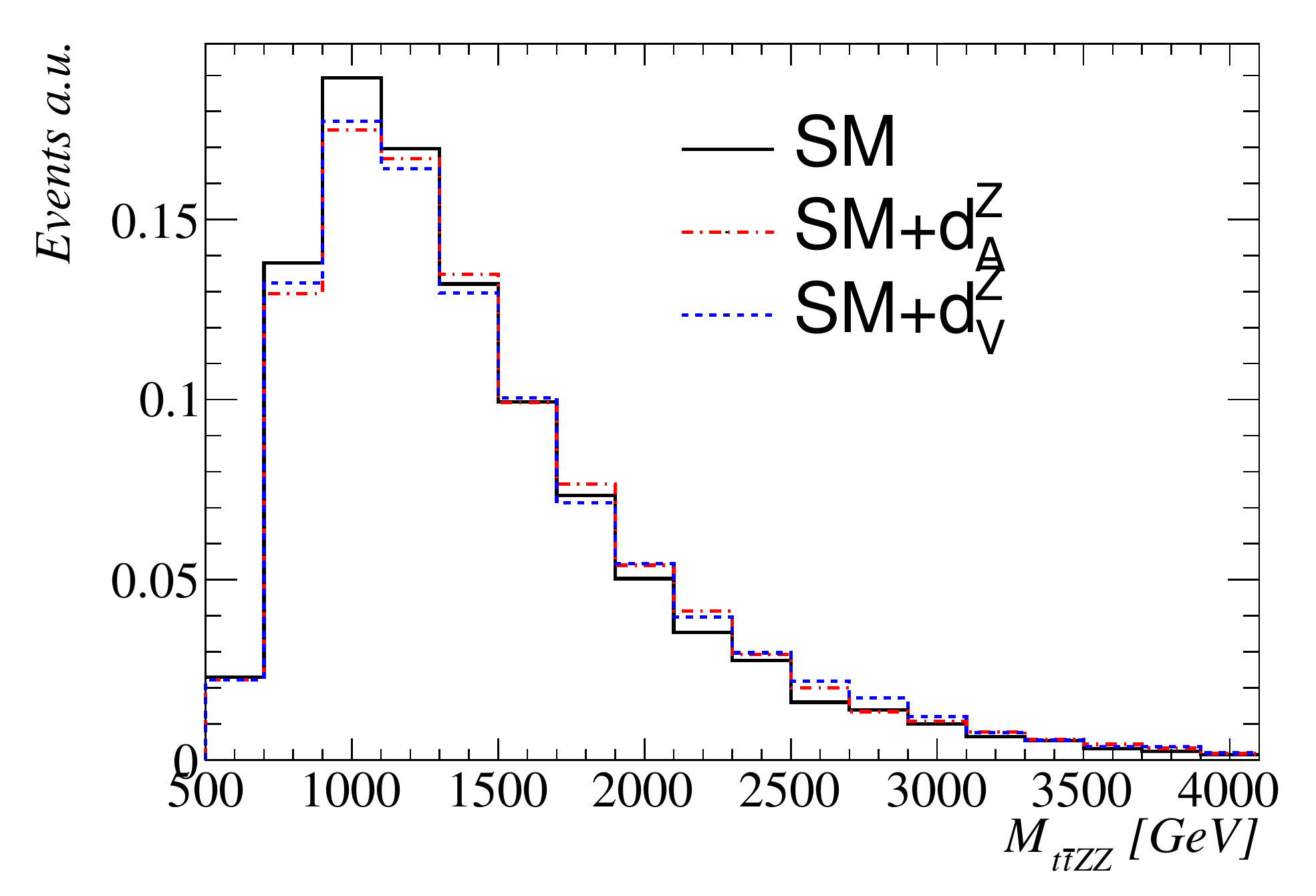}}  
\resizebox{0.4\textwidth}{!}{\includegraphics{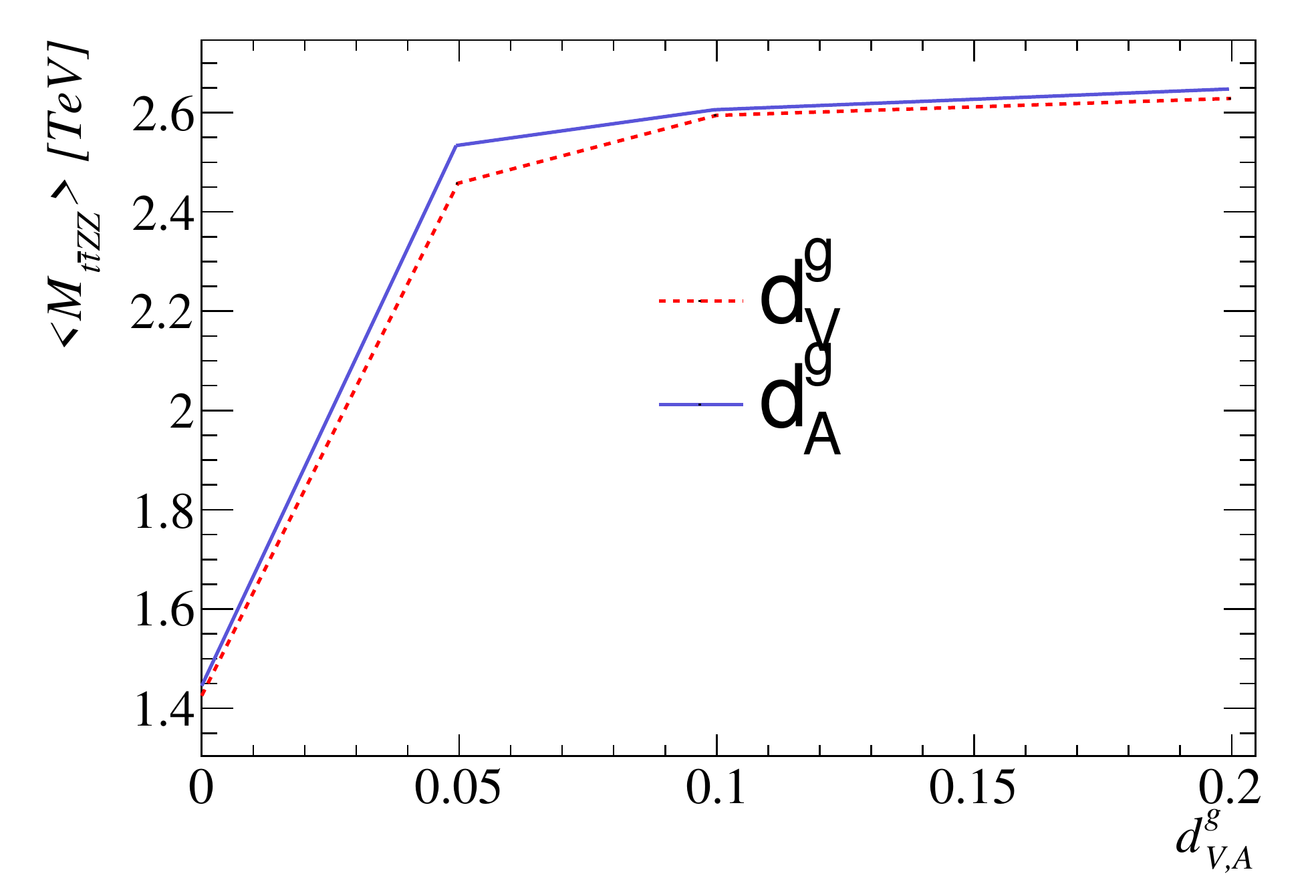}}  
\caption{  Top: The invariant mass distributions of the $t\bar{t}ZZ$ system for the SM and  $d^{g}_{V,A} = 0.05$ (left) and $d^{Z}_{V,A} = 0.05$ (right).
Bottom:  The functionality of $<M_{t\bar{t}ZZ}>$ versus $d^{g}_{V}$ and  $d^{g}_{A}$.}\label{m4p}
\end{center}
\end{figure}

In this section, We examine the information
 that could be derived from measuring of the  invariant
mass of  the system, i.e. $M_{t\bar{t}ZZ}$.
In the left side of Fig.\ref{m4p}, we display the normalized distributions
of $M_{t\bar{t}ZZ}$ for the SM and  two cases of $d^{g}_{V} = 0.05$
and $d^{g}_{A} = 0.05$.
One can see that the distribution of $M_{t\bar{t}ZZ}$ 
is peaked towards small masses for the SM case. While, in the presence of 
either $d^{g}_{V}$ or $d^{g}_{A}$ the peak substantially moves toward large values.
This could be traced back to the momentum dependence of these couplings.
In the bottom plot of Fig.~\ref{m4p}, the average value of $M_{t\bar{t}ZZ}$  distribution
is presented in terms of $d^{g}_{V}$ and $d^{g}_{A}$. As it can be seen, $<M_{t\bar{t}ZZ}>$
starts from around 1.42 TeV for the SM and grows significantly with $d^{g}_{V,A}$ coulings
and reaches up to around 2.45 TeV for $d^{g}_{V} = 0.05$ and 2.59 TeV for $d^{g}_{A} = 0.05$.
We see an explicit rapid raise in the $<M_{t\bar{t}ZZ}>$ with a small change in either $d^{g}_{V}$ or $d^{g}_{A}$
that certainly allows us to deeply probe the $gt\bar{t}$ structure. We note that the difference between  $<M_{t\bar{t}ZZ}>$
for the cases of $d^{g}_{V} = 0.05$ and $d^{g}_{A} = 0.05$ is around 150 GeV. 
The right panel of Fig.\ref{m4p} shows again the invariant mass distribution of the 
$t\bar{t}ZZ$ system for the SM and  two cases of $d^{Z}_{V} = 0.05$
and $d^{Z}_{A} = 0.05$. As it can be seen, the distributions are similar to the SM and 
almost we see no considerable change in the peak positions. 
The $<M_{t\bar{t}ZZ}>$  value for the cases of $d^{Z}_{V} = 0.05$ and  $d^{Z}_{A} = 0.05$
are 1.47 TeV and 1.48 TeV, respectively. The growth with respect to the SM value is at the level 
of 50 to 60 GeV which is not comparable with the one received in cases of $d^{g}_{V}$ or $d^{g}_{A}$.

The presented distributions are idealized, as 
no effects of parton showering, hadronization, object identification and  reconstruction, etc. are included.
Also,  the selection cuts and  background contamination are not considered.

\subsection{Angular observables}

\begin{figure}[htb]
\begin{center}
\vspace{1cm}
\resizebox{0.4\textwidth}{!}{\includegraphics{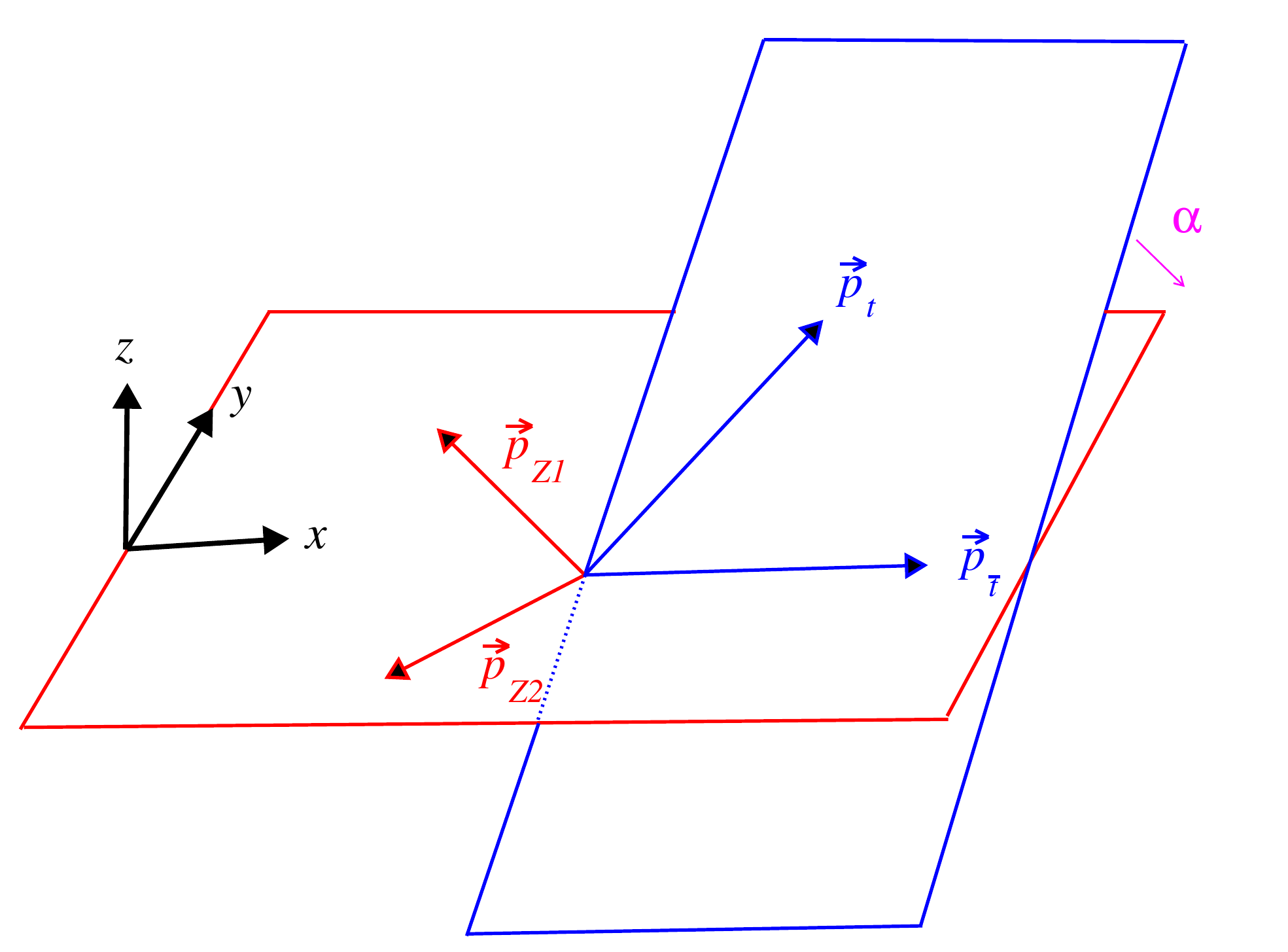}}  
\caption{  The angle $\alpha$ is shown which is the angle 
between two planes of $(\vec{p}_{t}~, ~\vec{p}_{\bar{t}})$ and $(\vec{p}_{Z_{1}}~,~\vec{p}_{Z_{2}})$. }\label{xzy}
\end{center}
\end{figure}

We now turn to the angular distributions.
Assuming the ability of the discrimination of  the top quarks and the $Z$ bosons
directions, we define the following observables:
\begin{eqnarray}\label{anggle}
z_{1} = \frac{(\vec{p}_{t}\times \vec{p}_{\bar{t}}).(\vec{p}_{Z_{1}}\times\vec{p}_{Z_{2}})}{|\vec{p}_{t}||\vec{p}_{\bar{t}}| |\vec{p}_{Z_{1}}| |\vec{p}_{Z_{2}}|}~, ~
z_{2} = \frac{(\vec{p}_{t}\times \vec{p}_{Z_{1}}).(\vec{p}_{\bar{t}}\times\vec{p}_{Z_{2}})}{|\vec{p}_{t}||\vec{p}_{\bar{t}}| |\vec{p}_{Z_{1}}| |\vec{p}_{Z_{2}}|}.
\end{eqnarray}
where $\vec{p}_{t(\bar{t})}$ and $\vec{p}_{Z_{1,2}}$ are the three-momenta of the top (anti-top)
and $Z$ bosons. For instance, the first observable $z_{1}$ is equal to $\cos\alpha$ where $\alpha$ is the angle 
between two planes of $(\vec{p}_{t}~, ~\vec{p}_{\bar{t}})$ and $(\vec{p}_{Z_{1}}~,~\vec{p}_{Z_{2}})$ as shown in Fig.~\ref{xzy}.

\begin{figure}[htb]
\begin{center}
\vspace{1cm}
\resizebox{0.4\textwidth}{!}{\includegraphics{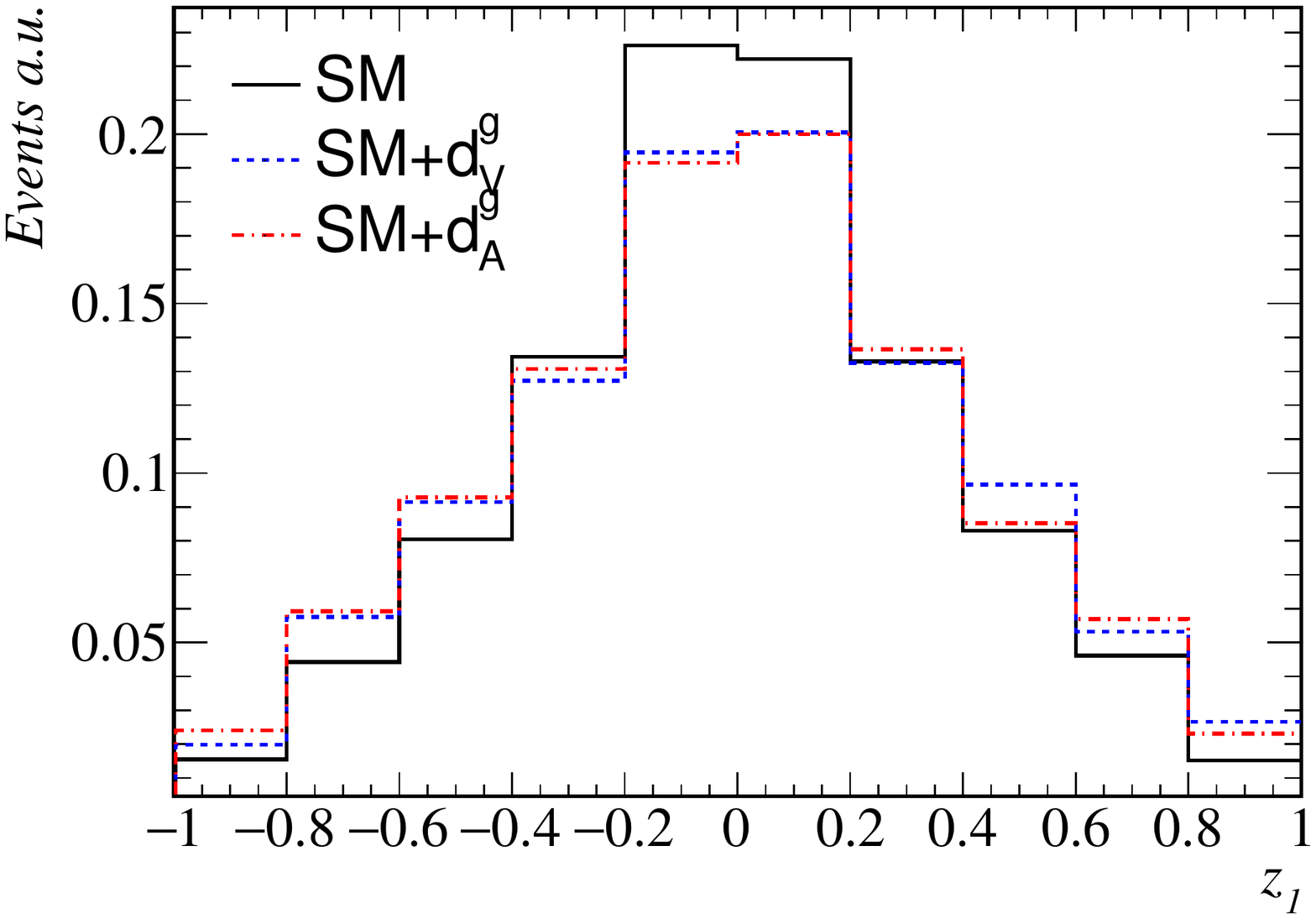}}  
\resizebox{0.4\textwidth}{!}{\includegraphics{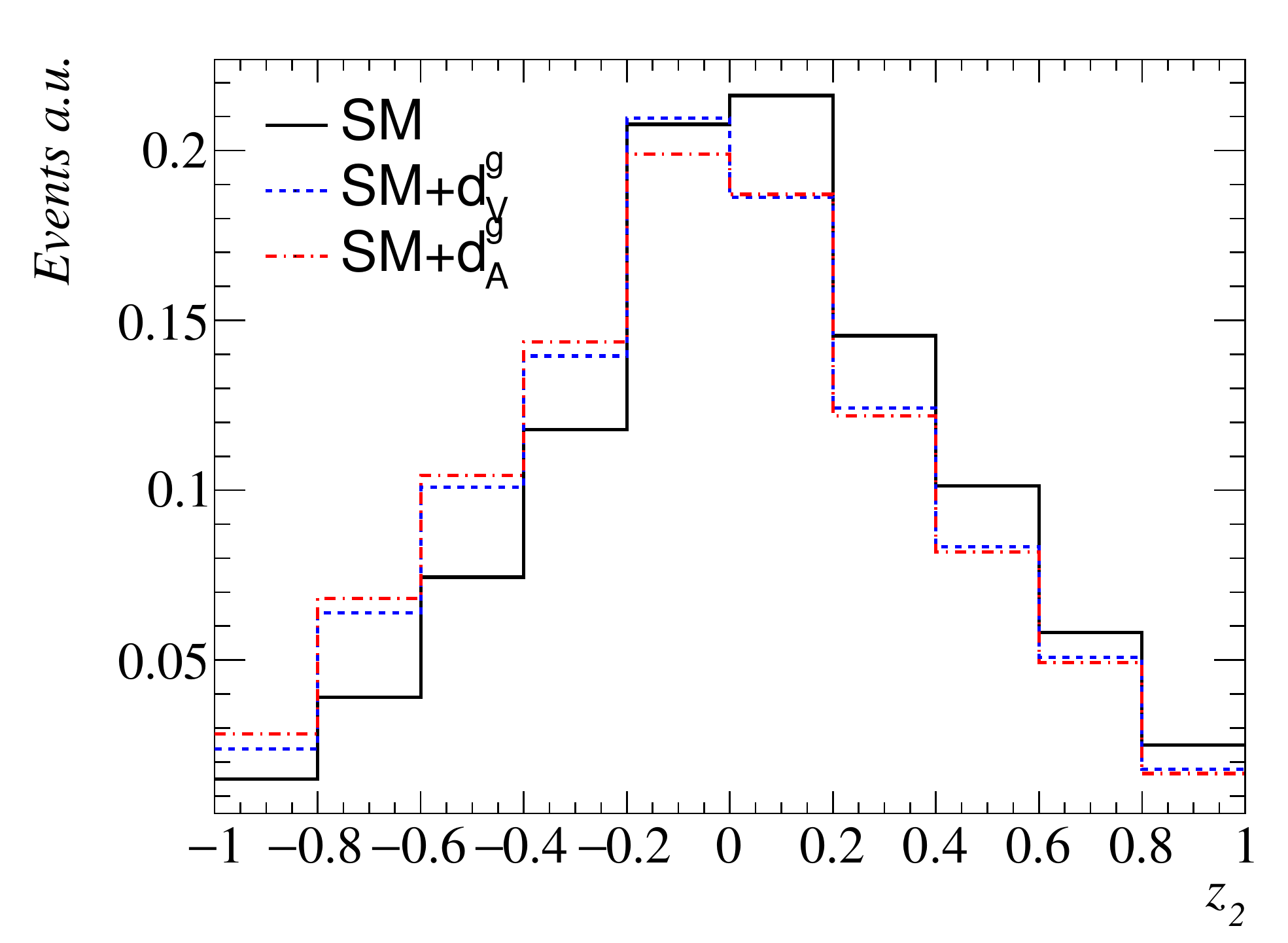}}  
\caption{  The $z_{1}$ and $z_{2}$ distributions for the SM, $d^{g}_{V} = 0.05$
and  $d^{g}_{A} = 0.05$ cases in the $pp\rightarrow t\bar{t}ZZ$ process.}\label{thth}
\end{center}
\end{figure}

In the definition of observable $z_{2}$, the first cross in the nominator is between the top quark and the $Z$ boson with highest $p_{T}$
and the second cross is between the anti-top quark and the other $Z$ boson.
For illustration, the shapes of $z_{1}$ and $z_{2}$ for the SM case (all couplings are set to zero)
and for instance for  $d^{g}_{V} = 0.05$  and $d^{g}_{A} = 0.05$ are depicted in Fig.\ref{thth}. 
As it can be seen, switching on the strong top quark dipole moments leads the
the $z_{1}$ and $z_{2}$ distributions to become wider and to have less peaked behavior with respect to the SM case.
A measure which reflects the heaviness of the tail of the $z_{1,2}$ distributions 
in the presence of $d^{g}_{V}$ and $d^{g}_{A}$ couplings is \textit{kurtosis}. 
The size of kurtosis for the SM distribution for $z_{1}$ ($z_{2}$) is found to be 0.172 (0.270) while
for $d^{g}_{V} = 0.05$ and $d^{g}_{A} = 0.05$, we find 
0.430 (0.431) and 0.432 (0.465), respectively. 
As expected, the kurtosis grows as the tail becomes heavier. 
It is notable that $z_{1}$ receives more change in kurtosis with respect to the SM case than $z_{2}$.

From the $z_{2}$ distributions, we also see that the presence of 
$d^{g}_{V}$ and $d^{g}_{A}$ couplings would lead events to be more distributed 
to $z_{2} < 0.0$ region than $z_{2} > 0.0$ region. To quantify that we define an asymmetry as:
\begin{eqnarray}
A(z_{2}) = \frac{N_{\rm events}(z_{2} > 0) - N_{\rm events}(z_{2} < 0)} {N_{\rm events}(z_{2} > 0) + N_{\rm events}(z_{2} < 0)}.
\end{eqnarray}
The denominator is the total number of events.
The value of $A(z_{2})$ for the SM is found to be $10\%$ and for SM+$d^{g}_{V}$ ($d^{g}_{V} = 0.05$) is $-7.5\%$. 
For SM+$d^{g}_{A}$ ($d^{g}_{A} = 0.05$),  $A(z_{2})$ amounts to $-9.0\%$. As it can be seen,  switching on 
$d^{g}_{V}$ or $d^{g}_{A}$  leads to a migration of events from $z_{2} > 0.0$ to  $z_{2} < 0.0$ and consequently 
change the sign of asymmetry $A(z_{2})$  from a positive to negative values. As a result, the measurement of $A(z_{2})$ asymmetry 
will provide valuable information on the new physics effects.
We have examined the sensitivities of the above angular observables to $d^{Z}_{V}$ and $d^{Z}_{A}$ and
no serious distortion is observed.

%
%
\section{Summary and conclusions}\label{summary}

Rare SM processes involving  multi-top-quark and multi-gauge-boson final states at the 
LHC provide an exciting opportunity to search for new physics effects.
To assess those effects, searches could be performed 
using the effective field theory approach which could affect
both the total cross sections and the differential distributions. Particularly, the 
impacts would be expected to  be significantly visible in processes
containing heavy particles in their final states.
In this paper, for the first time, we study the strong and weak electric ($d^{g,Z}_{A}$) and magnetic ($d^{g,Z}_{V}$) dipole moments 
of the top quark through the $t\bar{t}WW$ and $t\bar{t}ZZ$ channels at the LHC14.
As the SM values for $d^{g,Z}_{V}$ and $d^{g,Z}_{A}$ are very small,
in case of facing a situation with  $d^{g,Z}_{V,A}$ large enough,  $t\bar{t}WW$ and $t\bar{t}ZZ$ channels
provide promising ways to observe the corresponding excess over the expectation of the SM.

Based on the top quarks, $W$ bosons and $Z$ bosons decays, 
various signatures are available from which we have concentrated
on the much cleaner same-sign dilepton 
and four-lepton topologies for $t\bar{t}WW$ and $t\bar{t}ZZ$ channels, respectively.
Therefore, we assume the signals considered here are
adequately distinguishable from the SM backgrounds 
and a comprehensive study with including the backgrounds 
and detector effects are left for a future work. 
We find constraints of  $-0.09 \leq d^{Z}_{V} \leq 0.08$,  
$|d^{Z}_{A}| \leq 0.08$ for the weak dipole moments
and $-0.005\leq d^{g}_{V} \leq 0.006$,  $|d^{g}_{A}| \leq 0.005$ for the
strong dipole moments using 3 ab$^{-1}$ of the integrated luminosity 
of data. The results are comparable with the prospective ones reachable from 
$t\bar{t}$  and $t\bar{t}Z$ at the LHC.  However, 
there are rooms for significant improvements of the bounds 
which could be achieved by including different topologies 
and by taking into account the higher order QCD corrections
to signal processes.
Going beyond the total production cross section,  new angular observables are proposed to
probe the effects top quark dipole moments. We have found that the 
presence of top quark dipole moments could affect the final state angular configuration.
We  also show that the invariant mass distribution of $t\bar{t}ZZ$ system
is substantially sensitive to $d^{g}_{V}$ and $d^{g}_{A}$, pushing the peak to large mass region.
More investigation to examine the sensitivity to new couplings considering the impacts 
of final state reconstruction and detector
resolution are left to a future work.
 Another feature that is also  to be studied
 will be the disentanglement of 
CP-even and CP-odd operators through new observables.


\vspace{0.5cm}
%
{\bf Acknowledgments:}
M. Mohammadi Najafabadi would like to thank the Iran National Science Foundation
(INSF) for the financial support.
%

%
%


\begin{thebibliography}{99}

\bibitem{r0}
 M.~Cristinziani and M.~Mulders,
  J.\ Phys.\ G {\bf 44}, no. 6, 063001 (2017)
  doi:10.1088/1361-6471/44/6/063001
  [arXiv:1606.00327 [hep-ex]];   A.~Giammanco and R.~Schwienhorst,
  arXiv:1710.10699 [hep-ex];   A.~Giammanco, 
  Rev.\ Phys.\  {\bf 1}, 1 (2016)
  doi:10.1016/j.revip.2015.12.001
  [arXiv:1511.06748 [hep-ex]].
  
  \bibitem{r1}
 W.~Buchmuller and D.~Wyler,
  Nucl.\ Phys.\ B {\bf 268}, 621 (1986).
  doi:10.1016/0550-3213(86)90262-2

\bibitem{r2}
 J.~A.~Aguilar-Saavedra,
  Nucl.\ Phys.\ B {\bf 812}, 181 (2009)
  doi:10.1016/j.nuclphysb.2008.12.012
  [arXiv:0811.3842 [hep-ph]].
\bibitem{r3}
 B.~Grzadkowski, M.~Iskrzynski, M.~Misiak and J.~Rosiek,
  JHEP {\bf 1010}, 085 (2010)
  doi:10.1007/JHEP10(2010)085
  [arXiv:1008.4884 [hep-ph]].

\bibitem{r4}
 C.~Arzt, M.~B.~Einhorn and J.~Wudka,
  Nucl.\ Phys.\ B {\bf 433}, 41 (1995)
  doi:10.1016/0550-3213(94)00336-D
  [hep-ph/9405214].

\bibitem{r5}
 R.~Contino, A.~Falkowski, F.~Goertz, C.~Grojean and F.~Riva,
  JHEP {\bf 1607}, 144 (2016)
  doi:10.1007/JHEP07(2016)144
  [arXiv:1604.06444 [hep-ph]].


\bibitem{r6}
 D.~Atwood, S.~Bar-Shalom, G.~Eilam and A.~Soni,
  Phys.\ Rept.\  {\bf 347}, 1 (2001)
  doi:10.1016/S0370-1573(00)00112-5
  [hep-ph/0006032].


\bibitem{r7}
 K.~Whisnant, B.~L.~Young and X.~Zhang,
  Phys.\ Rev.\ D {\bf 52}, 3115 (1995)
  doi:10.1103/PhysRevD.52.3115
  [hep-ph/9410369];   K.~Fuyuto, W.~S.~Hou and E.~Senaha,
  arXiv:1705.05034 [hep-ph];  J.~M.~Cline, K.~Kainulainen and M.~Trott,
  JHEP {\bf 1111}, 089 (2011)
  doi:10.1007/JHEP11(2011)089
  [arXiv:1107.3559 [hep-ph]]; 
   M.~Jiang, L.~Bian, W.~Huang and J.~Shu,
  Phys.\ Rev.\ D {\bf 93}, no. 6, 065032 (2016)
  doi:10.1103/PhysRevD.93.065032
  [arXiv:1502.07574 [hep-ph]];
   A.~Kobakhidze, L.~Wu and J.~Yue,
  JHEP {\bf 1604}, 011 (2016)
  doi:10.1007/JHEP04(2016)011
  [arXiv:1512.08922 [hep-ph]].

      
\bibitem{rsavi}
 J.~A.~Aguilar-Saavedra, B.~Fuks and M.~L.~Mangano,
  Phys.\ Rev.\ D {\bf 91}, 094021 (2015)
  doi:10.1103/PhysRevD.91.094021
  [arXiv:1412.6654 [hep-ph]].
\bibitem{r8}
 W.~Bernreuther, L.~Chen, I.~García, M.~Perelló, Poeschl R., F.~Richard, E.~Ros and M.~Vos,
  arXiv:1710.06737 [hep-ex].


\bibitem{r9}
  S.~D.~Rindani, P.~Sharma and A.~W.~Thomas,
  JHEP {\bf 1510}, 180 (2015)
  doi:10.1007/JHEP10(2015)180
  [arXiv:1507.08385 [hep-ph]].


\bibitem{r10}
 Z.~Hioki and K.~Ohkuma,
  Phys.\ Rev.\ D {\bf 88}, 017503 (2013)
  doi:10.1103/PhysRevD.88.017503
  [arXiv:1306.5387 [hep-ph]].


\bibitem{r11}
 S.~S.~Biswal, S.~D.~Rindani and P.~Sharma,
  Phys.\ Rev.\ D {\bf 88}, 074018 (2013)
  doi:10.1103/PhysRevD.88.074018
  [arXiv:1211.4075 [hep-ph]].

\bibitem{r12}
 D.~Choudhury and P.~Saha,
  JHEP {\bf 1208}, 144 (2012)
  doi:10.1007/JHEP08(2012)144
  [arXiv:1201.4130 [hep-ph]].
  
\bibitem{r13}
 Z.~Hioki and K.~Ohkuma,
  Phys.\ Lett.\ B {\bf 716}, 310 (2012)
  doi:10.1016/j.physletb.2012.08.035
  [arXiv:1206.2413 [hep-ph]].


\bibitem{r14}
 W.~Bernreuther, D.~Heisler and Z.~G.~Si,
  JHEP {\bf 1512}, 026 (2015)
  doi:10.1007/JHEP12(2015)026
  [arXiv:1508.05271 [hep-ph]].


\bibitem{r15}
 S.~K.~Gupta, A.~S.~Mete and G.~Valencia,
  Phys.\ Rev.\ D {\bf 80}, 034013 (2009)
  doi:10.1103/PhysRevD.80.034013
  [arXiv:0905.1074 [hep-ph]].


\bibitem{r16}
 H.~Hesari and M.~Mohammadi Najafabadi,
  Phys.\ Rev.\ D {\bf 91}, no. 5, 057502 (2015)
  doi:10.1103/PhysRevD.91.057502
  [arXiv:1407.5887 [hep-ph]].

\bibitem{r17}
 S.~Yaser Ayazi, H.~Hesari and M.~Mohammadi Najafabadi,
  Phys.\ Lett.\ B {\bf 727}, 199 (2013)
  doi:10.1016/j.physletb.2013.10.025
  [arXiv:1307.1846 [hep-ph]].


\bibitem{r18}
 H.~Hesari and M.~Mohammadi Najafabadi,
  Mod.\ Phys.\ Lett.\ A {\bf 28}, 1350170 (2013)
  doi:10.1142/S0217732313501708
  [arXiv:1207.0339 [hep-ph]].

\bibitem{rcms}
  CMS Collaboration [CMS Collaboration],
  CMS-PAS-TOP-14-005.


\bibitem{r19}
  H.~Khanpour, S.~Khatibi, M.~Khatiri Yanehsari and M.~Mohammadi Najafabadi,
  Phys.\ Lett.\ B {\bf 775}, 25 (2017)
  doi:10.1016/j.physletb.2017.10.047
  [arXiv:1408.2090 [hep-ph]].


\bibitem{r20}
 S.~M.~Etesami, S.~Khatibi and M.~Mohammadi Najafabadi,
  Eur.\ Phys.\ J.\ C {\bf 76}, no. 10, 533 (2016)
  doi:10.1140/epjc/s10052-016-4376-2
  [arXiv:1606.02178 [hep-ph]].

\bibitem{r21}
 S.~Khatibi and M.~Mohammadi Najafabadi,
  Phys.\ Rev.\ D {\bf 90}, no. 7, 074014 (2014)
  doi:10.1103/PhysRevD.90.074014
  [arXiv:1409.6553 [hep-ph]].

\bibitem{r22}
W.~Bernreuther and Z.~G.~Si,
  Nucl.\ Phys.\ B {\bf 837} (2010) 90
  doi:10.1016/j.nuclphysb.2010.05.001
  [arXiv:1003.3926 [hep-ph]].


\bibitem{r23}
 M.~Schulze and Y.~Soreq,
  Eur.\ Phys.\ J.\ C {\bf 76}, no. 8, 466 (2016)
  doi:10.1140/epjc/s10052-016-4263-x
  [arXiv:1603.08911 [hep-ph]].

\bibitem{r24}
  R.~Rontsch and M.~Schulze,
  JHEP {\bf 1407}, 091 (2014)
  Erratum: [JHEP {\bf 1509}, 132 (2015)]
  doi:10.1007/JHEP09(2015)132, 10.1007/JHEP07(2014)091
  [arXiv:1404.1005 [hep-ph]].

\bibitem{r25}
 R.~Rontsch and M.~Schulze,
  JHEP {\bf 1508}, 044 (2015)
  doi:10.1007/JHEP08(2015)044
  [arXiv:1501.05939 [hep-ph]].


\bibitem{r26}
 S.~Fayazbakhsh, S.~T.~Monfared and M.~Mohammadi Najafabadi,
  Phys.\ Rev.\ D {\bf 92}, no. 1, 014006 (2015)
  doi:10.1103/PhysRevD.92.014006
  [arXiv:1504.06695 [hep-ph]].


\bibitem{r27}
 J.~F.~Kamenik, J.~Shu and J.~Zupan,
  Eur.\ Phys.\ J.\ C {\bf 72}, 2102 (2012)
  doi:10.1140/epjc/s10052-012-2102-2
  [arXiv:1107.5257 [hep-ph]].


\bibitem{r28}
 A.~O.~Bouzas and F.~Larios,
  Phys.\ Rev.\ D {\bf 88}, no. 9, 094007 (2013)
  doi:10.1103/PhysRevD.88.094007
  [arXiv:1308.5634 [hep-ph]].
\bibitem{r29}
Q.~H.~Cao, C.~R.~Chen, F.~Larios and C.-P.~Yuan,
  Phys.\ Rev.\ D {\bf 79}, 015004 (2009)
  doi:10.1103/PhysRevD.79.015004
  [arXiv:0801.2998 [hep-ph]].

\bibitem{r30}
  R.~Gaitan, E.~A.~Garces, J.~H.~M.~de Oca and R.~Martinez,
  Phys.\ Rev.\ D {\bf 92}, no. 9, 094025 (2015)
  doi:10.1103/PhysRevD.92.094025
  [arXiv:1505.04168 [hep-ph]].


\bibitem{r31}
M.~Casolino, T.~Farooque, A.~Juste, T.~Liu and M.~Spannowsky,
  Eur.\ Phys.\ J.\ C {\bf 75}, 498 (2015)
  doi:10.1140/epjc/s10052-015-3708-y
  [arXiv:1507.07004 [hep-ph]].


\bibitem{r32}
 C.~Englert, F.~Krauss, M.~Spannowsky and J.~Thompson,
  Phys.\ Lett.\ B {\bf 743}, 93 (2015)
  doi:10.1016/j.physletb.2015.02.041
  [arXiv:1409.8074 [hep-ph]].

\bibitem{r33}
 C.~Englert and M.~Spannowsky,
  Phys.\ Lett.\ B {\bf 740}, 8 (2015)
  doi:10.1016/j.physletb.2014.11.035
  [arXiv:1408.5147 [hep-ph]].


\bibitem{r34}
 C.~Englert, D.~Goncalves and M.~Spannowsky,
  Phys.\ Rev.\ D {\bf 89}, no. 7, 074038 (2014)
  doi:10.1103/PhysRevD.89.074038
  [arXiv:1401.1502 [hep-ph]].

\bibitem{r35}
 C.~Englert, A.~Freitas, M.~Spira and P.~M.~Zerwas,
  Phys.\ Lett.\ B {\bf 721}, 261 (2013)
  doi:10.1016/j.physletb.2013.03.017
  [arXiv:1210.2570 [hep-ph]].


\bibitem{r36}
 F.~Larios, T.~M.~P.~Tait and C.~P.~Yuan,
  Phys.\ Rev.\ D {\bf 57}, 3106 (1998)
  doi:10.1103/PhysRevD.57.3106
  [hep-ph/9709316].

\bibitem{r37}
 C.~Englert and M.~Russell,
  Eur.\ Phys.\ J.\ C {\bf 77}, no. 8, 535 (2017)
  doi:10.1140/epjc/s10052-017-5095-z
  [arXiv:1704.01782 [hep-ph]].

\bibitem{r38}
 J.~Chang, K.~Cheung, J.~S.~Lee and C.~T.~Lu,
  JHEP {\bf 1405}, 062 (2014)
  doi:10.1007/JHEP05(2014)062
  [arXiv:1403.2053 [hep-ph]].


\bibitem{r39}
 K.~m.~Cheung,
  Phys.\ Rev.\ D {\bf 53}, 3604 (1996)
  doi:10.1103/PhysRevD.53.3604
  [hep-ph/9511260].

\bibitem{r40}
  O.~Bessidskaia Bylund, F.~Maltoni, I.~Tsinikos, E.~Vryonidou and C.~Zhang,
  JHEP {\bf 1605}, 052 (2016)
  doi:10.1007/JHEP05(2016)052
  [arXiv:1601.08193 [hep-ph]].


\bibitem{r41}
 C.~Degrande, J.~M.~Gerard, C.~Grojean, F.~Maltoni and G.~Servant,
  JHEP {\bf 1103}, 125 (2011)
  doi:10.1007/JHEP03(2011)125
  [arXiv:1010.6304 [hep-ph]].

\bibitem{r42}
 D.~Buarque Franzosi and C.~Zhang,
  Phys.\ Rev.\ D {\bf 91}, no. 11, 114010 (2015)
  doi:10.1103/PhysRevD.91.114010
  [arXiv:1503.08841 [hep-ph]].

\bibitem{r43}
  D.~Barducci, M.~Fabbrichesi and A.~Tonero,
  Phys.\ Rev.\ D {\bf 96}, no. 7, 075022 (2017)
  doi:10.1103/PhysRevD.96.075022
  [arXiv:1704.05478 [hep-ph]].

\bibitem{r44}
  F.~Maltoni, E.~Vryonidou and C.~Zhang,
  JHEP {\bf 1610}, 123 (2016)
  doi:10.1007/JHEP10(2016)123
  [arXiv:1607.05330 [hep-ph]].


\bibitem{r45}
C.~Englert, L.~Moore, K.~Nordström and M.~Russell,
  Phys.\ Lett.\ B {\bf 763}, 9 (2016)
  doi:10.1016/j.physletb.2016.10.021
  [arXiv:1607.04304 [hep-ph]].
\bibitem{r46}
Y.~T.~Chien, V.~Cirigliano, W.~Dekens, J.~de Vries and E.~Mereghetti,
  JHEP {\bf 1602}, 011 (2016)
  [JHEP {\bf 1602}, 011 (2016)]
  doi:10.1007/JHEP02(2016)011
  [arXiv:1510.00725 [hep-ph]].


\bibitem{r47}
C.~Degrande, J.~M.~Gerard, C.~Grojean, F.~Maltoni and G.~Servant,
  JHEP {\bf 1207}, 036 (2012)
  Erratum: [JHEP {\bf 1303}, 032 (2013)]
  doi:10.1007/JHEP07(2012)036, 10.1007/JHEP03(2013)032
  [arXiv:1205.1065 [hep-ph]].

\bibitem{r48}
C.~Zhang and S.~Willenbrock,
  Phys.\ Rev.\ D {\bf 83}, 034006 (2011)
  doi:10.1103/PhysRevD.83.034006
  [arXiv:1008.3869 [hep-ph]].


\bibitem{r49}
 D.~Azevedo, A.~Onofre, F.~Filthaut and R.~Gonçalo,
  arXiv:1711.05292 [hep-ph].


\bibitem{r50}
 F.~Déliot, R.~Faria, M.~C.~N.~Fiolhais, P.~Lagarelhos, A.~Onofre, C.~M.~Pease and A.~Vasconcelos,
  arXiv:1711.04847 [hep-ph].


\bibitem{r51}
 J.~A.~Aguilar-Saavedra, M.~C.~N.~Fiolhais and A.~Onofre,
  JHEP {\bf 1207}, 180 (2012)
  doi:10.1007/JHEP07(2012)180
  [arXiv:1206.1033 [hep-ph]].


\bibitem{r52}
 M.~Mohammadi Najafabadi,
  JHEP {\bf 0803}, 024 (2008)
  doi:10.1088/1126-6708/2008/03/024
  [arXiv:0801.1939 [hep-ph]].


\bibitem{r53}
 H.~Hesari, H.~Khanpour and M.~Mohammadi Najafabadi,
  Phys.\ Rev.\ D {\bf 92}, no. 11, 113012 (2015)
  doi:10.1103/PhysRevD.92.113012
  [arXiv:1508.07579 [hep-ph]].


\bibitem{r54}
 E.~Boos, V.~Bunichev, L.~Dudko and M.~Perfilov,
  Int.\ J.\ Mod.\ Phys.\ A {\bf 32}, no. 02n03, 1750008 (2016)
  doi:10.1142/S0217751X17500087
  [arXiv:1607.00505 [hep-ph]].


\bibitem{r55}
 E.~Boos and L.~Dudko,
  Int.\ J.\ Mod.\ Phys.\ A {\bf 27}, 1230026 (2012)
  doi:10.1142/S0217751X12300268
  [arXiv:1211.7146 [hep-ph]].
  
\bibitem{r550}
  T.~G.~Rizzo,
  Phys.\ Rev.\ D {\bf 53} (1996) 6218
  doi:10.1103/PhysRevD.53.6218
  [hep-ph/9506351].


\bibitem{r56}
 P.~Torrielli,
  arXiv:1407.1623 [hep-ph].
\bibitem{r57}
 H.~van Deurzen, R.~Frederix, V.~Hirschi, G.~Luisoni, P.~Mastrolia and G.~Ossola,
  Eur.\ Phys.\ J.\ C {\bf 76}, no. 4, 221 (2016)
  doi:10.1140/epjc/s10052-016-4048-2
  [arXiv:1509.02077 [hep-ph]].

\bibitem{r58}
 E.~Alvarez, D.~A.~Faroughy, J.~F.~Kamenik, R.~Morales and A.~Szynkman,
  Nucl.\ Phys.\ B {\bf 915}, 19 (2017)
  doi:10.1016/j.nuclphysb.2016.11.024
  [arXiv:1611.05032 [hep-ph]].

\bibitem{r59}
 M.~Aaboud {\it et al.} [ATLAS Collaboration],
  Eur.\ Phys.\ J.\ C {\bf 77}, no. 1, 40 (2017)
  doi:10.1140/epjc/s10052-016-4574-y
  [arXiv:1609.01599 [hep-ex]].


\bibitem{r60}
A.~M.~Sirunyan {\it et al.} [CMS Collaboration],
  arXiv:1711.02547 [hep-ex].


\bibitem{r61}
 R.~Martinez and J.~A.~Rodriguez,
  Phys.\ Rev.\ D {\bf 65}, 057301 (2002)
  doi:10.1103/PhysRevD.65.057301
  [hep-ph/0109109].

\bibitem{r62}
R.~Martinez, M.~A.~Perez and N.~Poveda,
  Eur.\ Phys.\ J.\ C {\bf 53}, 221 (2008)
  doi:10.1140/epjc/s10052-007-0457-6
  [hep-ph/0701098].

\bibitem{nedm}
 J.~F.~Kamenik, M.~Papucci and A.~Weiler,
  Phys.\ Rev.\ D {\bf 85}, 071501 (2012)
  Erratum: [Phys.\ Rev.\ D {\bf 88}, no. 3, 039903 (2013)]
  doi:10.1103/PhysRevD.88.039903, 10.1103/PhysRevD.85.071501
  [arXiv:1107.3143 [hep-ph]].


\bibitem{r63}
 W.~Hollik, J.~I.~Illana, S.~Rigolin, C.~Schappacher and D.~Stockinger,
  Nucl.\ Phys.\ B {\bf 551}, 3 (1999)
  Erratum: [Nucl.\ Phys.\ B {\bf 557}, 407 (1999)]
  doi:10.1016/S0550-3213(99)00396-X, 10.1016/S0550-3213(99)00201-1
  [hep-ph/9812298].


\bibitem{r64}
J.~Bernabeu, D.~Comelli, L.~Lavoura and J.~P.~Silva,
  Phys.\ Rev.\ D {\bf 53}, 5222 (1996)
  doi:10.1103/PhysRevD.53.5222
  [hep-ph/9509416].
  
\bibitem{r65}
 T.~Ibrahim and P.~Nath,
  Phys.\ Rev.\ D {\bf 82}, 055001 (2010)
  doi:10.1103/PhysRevD.82.055001
  [arXiv:1007.0432 [hep-ph]].


\bibitem{r66}
  K.~Agashe, G.~Perez and A.~Soni,
  Phys.\ Rev.\ D {\bf 75}, 015002 (2007)
  doi:10.1103/PhysRevD.75.015002
  [hep-ph/0606293].

\bibitem{r67}
 C.~Englert and M.~Russell,
  Eur.\ Phys.\ J.\ C {\bf 77}, no. 8, 535 (2017)
  doi:10.1140/epjc/s10052-017-5095-z
  [arXiv:1704.01782 [hep-ph]].


\bibitem{mg5}
 J.~Alwall {\it et al.},
  JHEP {\bf 1407}, 079 (2014)
  doi:10.1007/JHEP07(2014)079
  [arXiv:1405.0301 [hep-ph]].

\bibitem{nnpdf}
  R.~D.~Ball {\it et al.} [NNPDF Collaboration],
  JHEP {\bf 1504}, 040 (2015)
  doi:10.1007/JHEP04(2015)040
  [arXiv:1410.8849 [hep-ph]].
\bibitem{feyn}
 A.~Alloul, N.~D.~Christensen, C.~Degrande, C.~Duhr and B.~Fuks,
  Comput.\ Phys.\ Commun.\  {\bf 185}, 2250 (2014)
  doi:10.1016/j.cpc.2014.04.012
  [arXiv:1310.1921 [hep-ph]].

\bibitem{ufo}
 C.~Degrande, C.~Duhr, B.~Fuks, D.~Grellscheid, O.~Mattelaer and T.~Reiter,
  Comput.\ Phys.\ Commun.\  {\bf 183}, 1201 (2012)
  doi:10.1016/j.cpc.2012.01.022
  [arXiv:1108.2040 [hep-ph]].


\bibitem{pythia8}
  T.~Sjostrand, S.~Mrenna and P.~Z.~Skands,
  Comput.\ Phys.\ Commun.\  {\bf 178}, 852 (2008)
  doi:10.1016/j.cpc.2008.01.036
  [arXiv:0710.3820 [hep-ph]].


\bibitem{jet}
 M.~Cacciari, G.~P.~Salam and G.~Soyez,
  JHEP {\bf 0804}, 063 (2008)
  doi:10.1088/1126-6708/2008/04/063
  [arXiv:0802.1189 [hep-ph]].

\bibitem{r80}
  F.~Maltoni, D.~Pagani and I.~Tsinikos,
  JHEP {\bf 1602}, 113 (2016)
  doi:10.1007/JHEP02(2016)113
  [arXiv:1507.05640 [hep-ph]].

\bibitem{bbb}
 B.~Grzadkowski and M.~Misiak,
  Phys.\ Rev.\ D {\bf 78}, 077501 (2008)
  Erratum: [Phys.\ Rev.\ D {\bf 84}, 059903 (2011)]
  doi:10.1103/PhysRevD.84.059903, 10.1103/PhysRevD.78.077501
  [arXiv:0802.1413 [hep-ph]].

\bibitem{ccc}
 C.~Bernardo, N.~F.~Castro, M.~C.~N.~Fiolhais, H.~Gonçalves, A.~G.~C.~Guerra, M.~Oliveira and A.~Onofre,
  Phys.\ Rev.\ D {\bf 90}, no. 11, 113007 (2014)
  doi:10.1103/PhysRevD.90.113007
  [arXiv:1408.7063 [hep-ph]].



\end{thebibliography}
\end{document}